\documentclass[longauth]{aa}

\usepackage{textcomp}
\usepackage[inline]{enumitem}
\usepackage{graphicx}
\usepackage[utf8]{inputenc}
\usepackage{physics}
\usepackage{txfonts}
\usepackage{etoolbox}
\usepackage{siunitx}

\usepackage[]{hyperref}
\DeclareUnicodeCharacter{2010}{-}

\begin{document}

\title{Study of the GeV to TeV morphology of the \texorpdfstring{$\gamma$}{gamma}-Cygni SNR (G\,78.2+2.1) with MAGIC and {\it Fermi}-LAT}

\titlerunning{Study of the \texorpdfstring{$\gamma$}{gamma}-Cygni SNR with MAGIC and {\it Fermi}-LAT}

\subtitle{Evidence for cosmic ray escape}

\author{
\small
MAGIC Collaboration:
V.~A.~Acciari\inst{1} \and
S.~Ansoldi\inst{2,24} \and
L.~A.~Antonelli\inst{3} \and
A.~Arbet Engels\inst{4} \and
D.~Baack\inst{5} \and
A.~Babi\'c\inst{6} \and
B.~Banerjee\inst{7} \and
U.~Barres de Almeida\inst{8} \and
J.~A.~Barrio\inst{9} \and
J.~Becerra Gonz\'alez\inst{1}
% ...
}

\institute { Inst. de Astrof\'isica de Canarias, E-38200 La Laguna, and Universidad de La Laguna, Dpto. Astrof\'isica, E-38206 La Laguna, Tenerife, Spain
\and Universit\`a di Udine, and INFN Trieste, I-33100 Udine, Italy
\and National Institute for Astrophysics (INAF), I-00136 Rome, Italy
\and ETH Zurich, CH-8093 Zurich, Switzerland
\and Technische Universit\"at Dortmund, D-44221 Dortmund, Germany
\and Croatian Consortium: University of Rijeka, Department of Physics, 51000 Rijeka; University of Split - FESB, 21000 Split; University of Zagreb - FER, 10000 Zagreb; University of Osijek, 31000 Osijek; Rudjer Boskovic Institute, 10000 Zagreb, Croatia
\and Saha Institute of Nuclear Physics, HBNI, 1/AF Bidhannagar, Salt Lake, Sector-1, Kolkata 700064, India
\and Centro Brasileiro de Pesquisas F\'isicas (CBPF), 22290-180 URCA, Rio de Janeiro (RJ), Brasil
\and IPARCOS Institute and EMFTEL Department, Universidad Complutense de Madrid, E-28040 Madrid, Spain
\and University of Lodz, Faculty of Physics and Applied Informatics, Department of Astrophysics, 90-236 Lodz, Poland
\and Universit\`a di Siena and INFN Pisa, I-53100 Siena, Italy
\and Deutsches Elektronen-Synchrotron (DESY), D-15738 Zeuthen, Germany
\and Istituto Nazionale Fisica Nucleare (INFN), 00044 Frascati (Roma), Italy
\and Max-Planck-Institut f\"ur Physik, D-80805 M\"unchen, Germany
\and Institut de F\'isica d'Altes Energies (IFAE), The Barcelona Institute of Science and Technology (BIST), E-08193 Bellaterra (Barcelona), Spain
\and Universit\`a di Padova and INFN, I-35131 Padova, Italy
\and Universit\`a di Pisa, and INFN Pisa, I-56126 Pisa, Italy
\and Universitat de Barcelona, ICCUB, IEEC-UB, E-08028 Barcelona, Spain
\and The Armenian Consortium: ICRANet-Armenia at NAS RA, A. Alikhanyan National Laboratory
\and Centro de Investigaciones Energ\'eticas, Medioambientales y Tecnol\'ogicas, E-28040 Madrid, Spain
\and Universit\"at W\"urzburg, D-97074 W\"urzburg, Germany
\and Finnish MAGIC Consortium: Finnish Centre of Astronomy with ESO (FINCA), University of Turku, FI-20014 Turku, Finland; Astronomy Research Unit, University of Oulu, FI-90014 Oulu, Finland
\and Departament de F\'isica, and CERES-IEEC, Universitat Aut\`onoma de Barcelona, E-08193 Bellaterra, Spain
\and Japanese MAGIC Consortium: ICRR, The University of Tokyo, 277-8582 Chiba, Japan; Department of Physics, Kyoto University, 606-8502 Kyoto, Japan; Tokai University, 259-1292 Kanagawa, Japan; RIKEN, 351-0198 Saitama, Japan
% ...
}

\date{Received XXX, 2020; accepted XXX, 2020}

\abstract
 {
 Diffusive shock acceleration (DSA) is the most promising mechanism to accelerate Galactic cosmic rays (CRs) in the shocks of supernova remnants (SNRs). It is based on particles scattering on turbulence ahead and behind the shock. The turbulence upstream is supposedly generated by the CRs, but this process is not well understood. The dominant mechanism may depend on the evolutionary state of the shock and can be studied via the CRs escaping upstream into the interstellar medium (ISM).
 }
 {
 Previous observations of the $\gamma$-Cygni SNR showed a difference in morphology between GeV and TeV energies. Since this SNR has the right age and is at the evolutionary stage for a significant fraction of CRs to escape, we aim to understand $\gamma$-ray emission in the vicinity of the $\gamma$-Cygni SNR. 
 }
 {
 We observed the region of the $\gamma$-Cygni SNR with the MAGIC Imaging Atmospheric Cherenkov telescopes between May 2015 and September 2017 recording 87 h of good-quality data. Additionally we analysed {\it Fermi}-LAT data to study the energy dependence of the morphology as well as the energy spectrum in the GeV to TeV range. The energy spectra and morphology were compared against theoretical predictions, which include a detailed derivation of the CR escape process and their $\gamma$-ray generation. 
 }
 {
 The MAGIC and {\it Fermi}-LAT data allowed us to identify three emission regions, which can be associated with the SNR and dominate at different energies. Our hadronic emission model accounts well for the morphology and energy spectrum of all source components. It constrains the time-dependence of the maximum energy of the CRs at the shock, the time-dependence of the level of turbulence, and the diffusion coefficient immediately outside the SNR shock. While in agreement with the standard picture of DSA, the time-dependence of the maximum energy was found to be steeper than predicted and the level of turbulence was found to change over the lifetime of the SNR.  
 }
 {} 

\maketitle 
   \section{Introduction}
\label{sec:sec_intro}

In the quest to identify the origin of Galactic cosmic rays (CRs), supernova remnants (SNRs) are the prime candidates. One main pillar of the so-called supernova paradigm is the mechanism of diffusive shock acceleration \citep[DSA; see][for a review]{Blasi2013, gabici_origin_2019}, which can efficiently transfer a fraction of the kinetic energy of the SNR shock wave to CRs. DSA predicts that CRs self-generate magnetic turbulence upstream of the shock that subsequently scatter them back downstream. For the most energetic CRs it is not clear whether such scattering centres are efficiently generated nor which is the principal mechanism responsible for their production. The answer probably depends on the evolutionary stage of the SNR. During the initial stage, when the shock speed is very high and the maximum energy of CRs is expected to increase, the non-resonant instability is thought to dominate \citep[see][for a review]{schure_diffusive_2012}. At later times, when the shock speed starts decreasing and the maximum energy should also decrease, the amplification is probably dominated by the resonant streaming instability. During both stages, at least a fraction of the highest energy CRs are expected to escape from the shock upstream \citep{ohira_escape_2010, malkov_analytic_2013, Celli+2019b, brose_acceleration_2020}. The process of DSA is inevitably connected to the escape of CRs into the interstellar medium (ISM). In contrast to the acceleration process, the mechanism of CR escape from the accelerator is not well understood, also due to the lack of clear observational signatures. Such evidence could be provided by $\gamma$-rays produced from interaction of escaping particles with the interstellar medium surrounding the SNR \citep{aharonian_emissivity_1996, Gabici+2009}. 

Because the amount of CRs escaping is small, young SNRs are not expected to show clear signatures of CR-ISM interaction. Recently the \citet{h.e.s.s._collaboration_h.e.s.s._2018} observed $\gamma$-rays beyond the X-ray emission region in the SNR RX\,J1713.7-3946. However, the angular resolution of the $\gamma$-ray telescopes was insufficient to distinguish whether the $\gamma$-rays are generated by CRs escaping the SNR or if they are the signature of the shock precursor\footnote{With ``{\it precursor}'' we refer to the region upstream of the shock where particles diffuse but are still bound to the shock.}.

During the adiabatic or Sedov-Taylor phase the shock velocity decreases significantly and large fractions of particles are released into the ISM. The interaction of such SNRs or their escaping CRs with dense molecular clouds was observed for a number of SNRs \citep[W28, IC443, W44, W51C; see][for a review]{slane_supernova_2015}. However, for such mature SNRs the shock has already encountered molecular clouds, and even low energy CRs have escaped from the accelerator, so the escape process cannot be studied in isolation.

With an age of ${\sim}7\times10^{3}$\,years, the $\gamma$-Cygni SNR is slightly older than RX\,J1713.7-3946, but younger than the middle-aged SNRs. Its circular radio shell suggests that, on the large scale, the hydrodynamic evolution has not been affected by density anisotropies so far and X-ray observations confirm that the SNR is clearly in its adiabatic phase; it is therefore an interesting target to search for signatures of escaping CRs.

\subsection{The \texorpdfstring{$\gamma$}{gamma}-Cygni SNR (G\,78.2+2.1)}

The $\gamma$-Cygni SNR (also called G\,78.2+2.1) is located in the heart of the Cygnus region close to the bright $\gamma$-Cygni star, Sadr \citep[mag=2.2, ][]{hoffleit_bright_1991}. Since it hosts the pulsar PSR 2021+4026, which is likely associated with the SNR \citep{hui_detailed_2015}, it is believed to be the debris of a core collapse supernova.

At radio wavelengths the SNR shows a distinctively circular shell with a diameter of ${\sim}\,1\degr$ \citep{higgs_true_1977, wendker_cygnus_1991, kothes_catalogue_2006}. The emission is brighter towards the south-east and north-west of the shell than along the north-east south-west axis. Further studies from \citet{zhang_multi-frequency_1997} and \citet{ladouceur_new_2008} found that the flux spectral index $\alpha_{\nu}$ varies between ${\sim}\,0.8$ and ${\lesssim}\,0.4$ across the SNR. The softest index is found in the bright south-eastern part, while the spectrum is harder ($\sim\,0.55$) in the north-west and south-west \citep{ladouceur_new_2008}. Based on radio observations and using $\Sigma$-D relations , HI velocity measurements, and association with the Wolf-Rayet binary V444 Cyg, the distance to $\gamma$-Cygni was determined to be between 1.5\,--\,2.6\,kpc (see Table \ref{table:snr_properties}). It is unclear whether the SNR is surrounded by a larger HI shell. \citet{gosachinskij_hi_2001} reported a shell of {2\fdg{}0}\,--\,{2\fdg{}8}\,$\times$\,{2\fdg{}5}\,--\,{3\fdg{}5} diameter centred approximately at the SNR, but noted that the HI structures are not necessarily at the same distance. \citet{ladouceur_new_2008} observed structures in emission bordering the SNR shell, but \citet{leahy_x-ray_2013} claimed that those structures are absorption features by layers situated in front of the SNR. Given its position in a OB region, it is plausible that the SNR might be surrounded by a HI cavity blown by the wind of the progenitor's stellar wind \citep{lozinskaya_supernova_2000}.

Observations of CO lines did not reveal any interaction of the SNR with molecular material \citep{higgs_co_1983} except for a hint at the south-eastern part \citep{fukui_co_1988}. The search for maser emission led to a negative result \citep{frail_survey_1996}.

No optical counterpart of the SNR has been detected. \citet{mavromatakis_deep_2003} searched for optical emission lines ([NII], [SII], and [OIII]) and found patchy emission towards the south, south-east, and north-west of the SNR. In the south-east hints for shock-heated gas suggest that the low density medium, in which the SNR evolves, contains clouds with pre-shock densities of ${\sim}$20\,cm$^{-3}$ and a shock velocity of $\sim750\,{\rm km\,s}^{-1}$. The author further inferred that most of the hot dust and absorbing matter lies in the foreground of the SNR, which possibly obscures most of the optical emission.

In the X-ray band, the emission is dominated by shock-heated gas as expected for a Sedov-Taylor (ST) phase SNR, even in the case of efficient particle acceleration \citep{castro_impact_2011}. The structure of the X-ray emission correlates with the radio band,  except it is also bright in the south-west. The post-shock gas temperature indicates a shock speed of ${\sim}\,1000$\,km\,s$^{-1}$ in all parts of the SNR \citep{higgs_x-ray_1983, lozinskaya_supernova_2000, uchiyama_asca_2002, leahy_x-ray_2013}. The gas temperature in the centre suggests that the reverse shock hit the centre about 1900\,yr ago and thus the SNR is fully adiabatic \citep{hui_detailed_2015}. Towards the northern shell the X-ray emission extends beyond the SNR radio shell and may partially be produced by the stellar wind in the foreground \citep{leahy_x-ray_2013}. Additionally, \citet{uchiyama_asca_2002} found three clumps of hard X-ray emission in the north-west, of which two are likely of Extragalactic origin \citep{leahy_x-ray_2013}. One (C2), however, may hint at bremsstrahlung clumps and a denser medium in that region. 

At GeV energies the {\it Fermi} Large Area Telescope ({\it Fermi}-LAT) observed extended emission over all of the SNR radio shell \citep{lande_search_2012}. The source was modelled with a disk of $0\fdg63\pm0\fdg05_{\rm stat}\pm0\fdg04_{\rm sys}$ radius centred on ($\alpha=305\fdg25$, $\delta=40\fdg52$; J2000) and a power-law spectrum with index $-2.42\pm0.19$. Later analyses of the SNR, however, found different extensions to better describe the data \citep[][]{acero_first_2016, ackermann_search_2017} and \citet{fraija_gigaelectronvolt_2016} found the spectral index in the north-west of the shell to be harder compared to the other parts. Furthermore, the {\it Fermi}-LAT collaboration discovered the $\gamma$-ray bright pulsar PSR\,J2021+4026, the only variable $\gamma$-ray pulsar known so far. It has a spin-down power of $\dot{E}_{\rm SD}{\sim}10^{35}$\,erg\,s$^{-1}$ and a characteristic age of $\tau_{\rm C}{\sim}77$\,kyr \citep{abdo_detection_2009}. The pulsar spectrum follows a power-law with exponential cutoff with a cutoff energy of $E_{\text{c}}=2.37\pm0.06$\,GeV \citep{allafort_psr_2013}.

At TeV energies the VERITAS telescopes discovered extended emission at the north-west of the shell, which was modelled with a Gaussian source (VER\,J2019+407) centred at ($\alpha=305\fdg02$, $\delta=40\fdg76$; J2000) \citet{aliu_discovery_2013}. The observation time was 21.4\,h. The spectrum of the extended emission followed a power-law (index of $-2.37$). In \citet{abeysekara_very_2018} VERITAS updated their findings with an increased exposure of ${\sim}40$\,h resulting in a softer spectral index ($-2.79\pm0.39_{\rm stat}\pm0.20_{\rm sys}$). The source is also listed in the 2nd HAWC catalogue as 2HWC\,J2020+403 \citep{abeysekara_2hwc_2017}. Its centre is at ($\alpha=305\fdg16$, $\delta=40\fdg37$; J2000) and the spectral index of the measured power-law is $-2.95\pm0.10$. However, the angular resolution of the HAWC detector was not sufficient to determine the size or to detect substructures. 

\begin{table*}
\caption{Physical parameters of the $\gamma$-Cygni SNR based on various measurements. The ranges only reflect the values as given in the corresponding publications.}             
\label{table:snr_properties}      
\centering                          
\linespread{1.15}\selectfont
\begin{tabular}{l c c c}        
\hline\hline                 
Characteristic & value used in this work & value range & References \\    
\hline                        
Radius [$^\circ$]    & 0.53 & 0.51\,--\,0.56   & \ref{item:higgs}, \ref{item:wendker}, \ref{item:kothes} \\
Distance [kpc]			         & 1.7  & 1.5\,--\,2.6   & \ref{item:higgs}, \ref{item:landecker}, \ref{item:lozinskaya}, \ref{item:uchiyama}, \ref{item:leahy}\\
Age [kyr]			             & 7    & 4\,--\,13      & \ref{item:higgs}, \ref{item:lozinskaya}, \ref{item:uchiyama}, \ref{item:leahy}\\
shock speed [km/s] 		         & 1000 & 600\,--\,1500  & \ref{item:higgs}. \ref{item:higgs2}, \ref{item:lozinskaya}, \ref{item:uchiyama}, \ref{item:leahy} \\
gas density at $\gamma$-Cygni [1/cm$^{3}$] 	 & 0.2  & 0.14\,--\,0.32 & \ref{item:saken}, \ref{item:lozinskaya}, \ref{item:leahy} \\
explosion energy [$10^{51}$ erg] & 1 	& 0.8\,--\,1.1   & \ref{item:mavromatakis}, \ref{item:leahy} \\
\hline                                   
\end{tabular}
\linespread{1.0}\selectfont
\tablebib{ \begin{enumerate*}[label={(\arabic*)}]
  \item \citet{higgs_true_1977}\label{item:higgs}; \item \citet{landecker_atomic_1980}\label{item:landecker}; \item \citet{higgs_x-ray_1983}\label{item:higgs2}; \item \citet{wendker_cygnus_1991}\label{item:wendker}; \item \citet{saken_iras_1992}\label{item:saken}; \item \citet{lozinskaya_supernova_2000}\label{item:lozinskaya}; \item \citet{uchiyama_asca_2002}\label{item:uchiyama}; \item\citet{mavromatakis_deep_2003}\label{item:mavromatakis}; \item \citet{kothes_catalogue_2006}\label{item:kothes}; \item \citet{leahy_x-ray_2013}\label{item:leahy}
  \end{enumerate*}
}
\end{table*} 
Table \ref{table:snr_properties} summarises the properties of the $\gamma$-Cygni SNR from the given references. In addition to the aforementioned characteristics, it lists the age, the density inside the shell, and the explosion energy. The age is inferred from the size of the radio shell, the shock speed, and the particle density assuming a Sedov-Taylor model (see e.g. Eqs. \eqref{eq:ST} and \eqref{eq:ST2}). Based on these measurements and estimates, in the following we will assume a distance of 1.7\,kpc, an age of 7000\,yrs, an explosion energy of 10$^{51}$\,ergs, and a shock speed of 10$^{3}$\,km\,s$^{-1}$. For the centre of the SNR radio shell we will use ($\alpha=305\fdg3$, $\delta=40\fdg43$; J2000). As the ejecta mass is unknown, but the SNR likely resulted from a core-collapse supernova of an OB star, we use a canonical value of $M_{\rm ej}=5\,{\rm M}_{\odot}$ for type II supernovae \citep{chevalier_interaction_1977}. It is important to note that the ranges given in Table \ref{table:snr_properties} just consider the optimal values from the listed publications excluding the uncertainties. The uncertainty ranges of all measurements are similar and thus the average values are still a reasonable representation. The parameters are further correlated due to their connection via the Sedov-Taylor model. Combining all estimates considering their statistical uncertainties, systematic uncertainties of each instrument and method, and their correlation is beyond the scope of this work. Accordingly, when using the extreme values from Table \ref{table:snr_properties} in our estimations later on, the resulting ranges are suggestive rather than accurate uncertainty intervals.

Overall the observed properties make the $\gamma$-Cygni SNR a prime example for a Sedov-Taylor phase SNR and for studying the possible escape of CRs. The discrepancy between the morphology at GeV energies observed by {\it Fermi}-LAT and the concentrated emission at TeV energies reported by VERITAS indeed suggests an ongoing, energy dependent process (see morphology described above or compare Figure 25 from \citet{lande_search_2012} and Figure 1 from \citet{aliu_discovery_2013}). We report on the observation of $\gamma$-Cygni with the MAGIC telescopes and combine them with an analysis of {\it Fermi}-LAT data to explore the discrepancy in GeV to TeV regime in greater detail.

   \section{Observations and data analysis}

\subsection{The MAGIC telescopes observations and data analysis}
\label{sec:MAGIC_ana}
The MAGIC (Major Atmospheric Gamma Imaging Cherenkov) telescopes are a system of two 17 m diameter imaging Cherenkov telescopes (IACTs) located at 2200\,m altitude above sea level at the Observatorio del Roque de los Muchachos on the Canary island La Palma, Spain $(28^{\circ}\,46^{\prime}\,\text{N},\,17 ^{\circ}\,53^{\prime}\,\text{W})$. The telescopes detect $\gamma$-ray induced extensive air showers in the atmosphere via their Cherenkov light. The telescopes are operated in stereoscopic mode, in which only showers triggering both telescopes are recorded. The telescopes cover the energy range from ${\sim}\,30$\,GeV to $E\,{>}\,100$\,TeV and have a field of view of $3\fdg5$ diameter. At low zenith angles $Zd\,{<}\,30\degr$ and within 50\,h, MAGIC can detect point sources above 200\,GeV at a flux level of $(0.66\,\pm\,0.03)$\% of the Crab Nebula flux; at medium zenith angles $30\degr\,{<}\,Zd\,{<}\,45\degr$ this level increases to $(0.76\,\pm\,0.04)$\% \citep{aleksic_major_2016}.\\

The observations for this work were performed over two periods between May and November 2015 and between April and September 2017. The data of the latter period include dim and moderate moon data classified according to \citet{ahnen_performance_2017}, though the sensitivity is comparable to the one under dark conditions. The data cover a zenith range from $10\degr$ to $55\degr$. The observations used two pointing positions in wobble mode with an offset of $0\fdg6$ from the VERITAS source location.  

Starlight increases the number of photoelectrons (ph.e.) in the pixels close to the position of Sadr in the camera. Hence, the pointing directions were chosen to have the same angular distance to Sadr and thereby to reduce a systematic mismatch between both pointings. Additionally, the position of Sadr was kept outside the trigger region of the camera (up to 1\fdg17 from the camera centre) so as not to increase the rate of spurious events. If the light yield in a camera pixel exceeds the safety threshold, it is switched off to save the photo-multiplier from ageing. This condition applied to 2\,--\,3 pixels at the position of Sadr neighboured by 12 pixels with higher light content. Artefacts from these features survived the analysis procedure at images sizes of $size\,{<}\,150$\,ph.e., where the additional star light or lost pixels significantly affected shower images arriving at the star's position. We thus apply a size cut of $size\,{>}\,$150\, ph.e. to the MAGIC data implying an energy threshold of $250$\,GeV. This limit was well above the energy threshold resulting from the general observational conditions (Moon conditions or zenith range). The MAGIC angular resolution, characterised by the point spread function (PSF), for this study was estimated to be $0\fdg08$ (68\% containment radius) at $E\,{>}\,250$\,GeV.

The MAGIC data were analysed using the MAGIC Analysis and Reconstruction Software \citep[MARS, ][]{zanin2013mars}. The analysis involved quality selection, cleaning of the shower images from night sky background, Hillas parametrisation, stereo reconstruction based on the disp method, and $\gamma$/hadron separation based on a random forest classifier. The data quality was controlled by monitoring the transparency of the atmosphere with a LIDAR system during observations \citep{fruck_novel_2014}. In this analysis we only included data with an atmospheric transmission above 85\,\% of the optimal transparency up to 12\,km above the telescopes. The cleaning levels were adapted depending on the night sky brightness following \citet{ahnen_performance_2017}. After quality cuts, the total dead-time corrected observation time amounted to 85\,h.  

We analysed the high level data with the SkyPrism spatial likelihood analysis package \citep{vovk_spatial_2018}. SkyPrism contains routines for computing the event count map, the background map, and the instrument response functions (PSF, energy migration matrix, and exposure map). Based on these, SkyPrism fits a user-defined source model to the measured event maps minimising the negative log-likelihood estimate. This way the package estimates detection flux normalisations and, if performed in several energy bins, extracts the source spectra for the considered model. Statistical inference in SkyPrism is based on a likelihood ratio (LRT) test defining the test statistic (TS)

\begin{equation}
\label{eq:LRT}
 TS = 2\left[ \ln\left(\mathcal{L}_{1}\right) - \ln\left(\mathcal{L}_{0}\right) \right]\,,
\end{equation}

\noindent where $\mathcal{L}_{0}$ is the likelihood value of the null hypothesis and $\mathcal{L}_{1}$ the one of the hypothesis being tested. The TS value can be converted to statistical significance via the Wilk's theorem \citep{wilks_large-sample_1938}. If the null hypothesis is true, TS follows a $\chi^2$ distribution with $n$ degrees of freedom, where $n$ is the number of additional parameters in the hypothesis model. However, when testing the presence of source component (i.e. when the LRT is performed on the boundary of the parameter space with a single additional free parameter, the flux of the new source component), the TS is converted to Gaussian significances by $S=\sqrt{TS}\sigma$ \citep{mattox_likelihood_1996, protassov_statistics_2002}. 

We used the "exclusion map" method for generating the background map excluding a circular region of 0\fdg56 deg around the radio centre ($\alpha=305\fdg3$, $\delta=40\fdg43$; J2000) and around the VERITAS centre ($\alpha=305\fdg02$, $\delta=40\fdg76$; J2000). The considered MAGIC region-of-interest (RoI), used in this work, had a size of $2\fdg5\times2\fdg5$ with a pixel size of $0\fdg02\times0\fdg02$. As the bins were smaller than the PSF, the spatial pixels were highly correlated, which is accounted for in the likelihood analysis. A description of the procedure used to estimated the systematic uncertainties on the source localisations and spectral parameters can be found in the appendix Sect. \ref{sec:Sys_Uncertainties}.

\subsection{The {\it Fermi}-LAT data observations and data analysis}
\label{sec:Fermi_ana}

The {\it Fermi}-LAT is a $\gamma$-ray telescope onboard the {\it Fermi} Gamma-ray Space Telescope \citep{atwood_large_2009}. Using the pair-conversion technique it is designed to observe the energy range between 20\,MeV and $E\,{>}$\,300\,GeV.

This study used data from ${\sim}\,9$ years of observation between 2008 Oct 27 and 2017 Sep 12 processed with the Pass 8\,R2 reconstruction \citep{atwood_pass_2013} as provided by the Fermi Science Support Center (FSSC). The data were analysed using the Fermi Science Tools (version v11r5p3\footnote{https://fermi.gsfc.nasa.gov/ssc/data/analysis/software}) in combination with the Fermipy package \citep[version 0.17.3, ][]{wood_fermipy_2017}. We chose the 'Source' selection cuts and instruments responses (P8R2\_SOURCE\_V6\footnote{https://fermi.gsfc.nasa.gov/ssc/data/Cicerone/\newline{}Cicerone\_Data\_Exploration/Data\_preparation.html}) for a balance between precision and photon count statistics. Further, we split the data according to the 4 PSF classes and combined them in a joint likelihood function. As for MAGIC, the {\it Fermi}-LAT analysis uses the LRT of Eq. \eqref{eq:LRT} for statistical inference. Accordingly, the TS approximates a $\chi^2$ distribution under the null hypothesis, except when testing the presence of a new component on the boundary, in which case $S=\sqrt{TS}\sigma$ applies as well.

The zenith angle was limited to 105\degr, a time filter was applied (DATA\_QUAL>0 \&\& LAT\_CONFIG==1), and the energy dispersion was considered for all sources except the Galactic and Extragalactic diffuse emission. The used interstellar emission model was 'gll\_iem\_v06.fits' and the isotropic diffuse emission model was 'iso\_P8R2\_SOURCE\_V6\_PSF[0,1,2,3]\_v06.txt' for the corresponding PSF class.

Below ${\sim}\,$10\,GeV the emission from the pulsar (PSR\,J2021+4026) dominates over the flux from the SNR. Using the off-pulse phase of the pulsar does not sufficiently suppress its contribution since the flux difference between the pulse peak and the off-pulse emission is only ${\sim}\,$30\,\% \citep{allafort_psr_2013}. To reliably disentangle both components, we limited the energy range of the {\it Fermi}-LAT data to 5\,GeV\,--\,500\,GeV. At $E\,{>}\,$5\,GeV the 95\,\% containment radius of the {\it Fermi}-LAT PSF is smaller than the radius of the SNR. We chose the RoI to be $10\degr\,\times\,10\degr$ around the radio centre of the SNR, a spatial bin-size of $0\fdg05$, and split the energy range in 18 bins (9/decade).

To model the contribution from background sources (including PSR\,J2021+4026) in the RoI, we used the FL8Y source list \footnote{\url{https://fermi.gsfc.nasa.gov/ssc/data/access/lat/fl8y/}} as a starting point considering sources within $15^{\circ}$ from the centre of the analysis. The source FL8Y\,J2021.0+4031e corresponding to the $\gamma$-Cygni SNR was removed from the model. After running the "optimize" procedure of Fermipy, we removed all sources with $TS\,{<}\,$16 or having no predicted counts. For the Cygnus Cocoon we used the spatial template of the extended source archive V18 provided online by the FSSC. A point source search in the RoI resulted in one significant, positive excess of ${>}\,5\,\sigma$ over our model coinciding with a 2\,FHL source (2FHL\,J2016.2+3713), which we added to our model. The remaining residuals stayed below ${<}\,5\,\sigma$ across the entire sky map. Sources related to the $\gamma$-Cygni SNR system are discussed in Sect. \ref{sec:Fermi_results}. The treatment of the systematic uncertainties on the source localisations and spectral parameters are described in the appendix Sect. \ref{sec:Sys_Uncertainties}.

   \section{Analysis results}

\subsection{MAGIC morphological results}
\label{sec:MAGIC_detections}

\begin{figure}
	\includegraphics[width=\columnwidth]{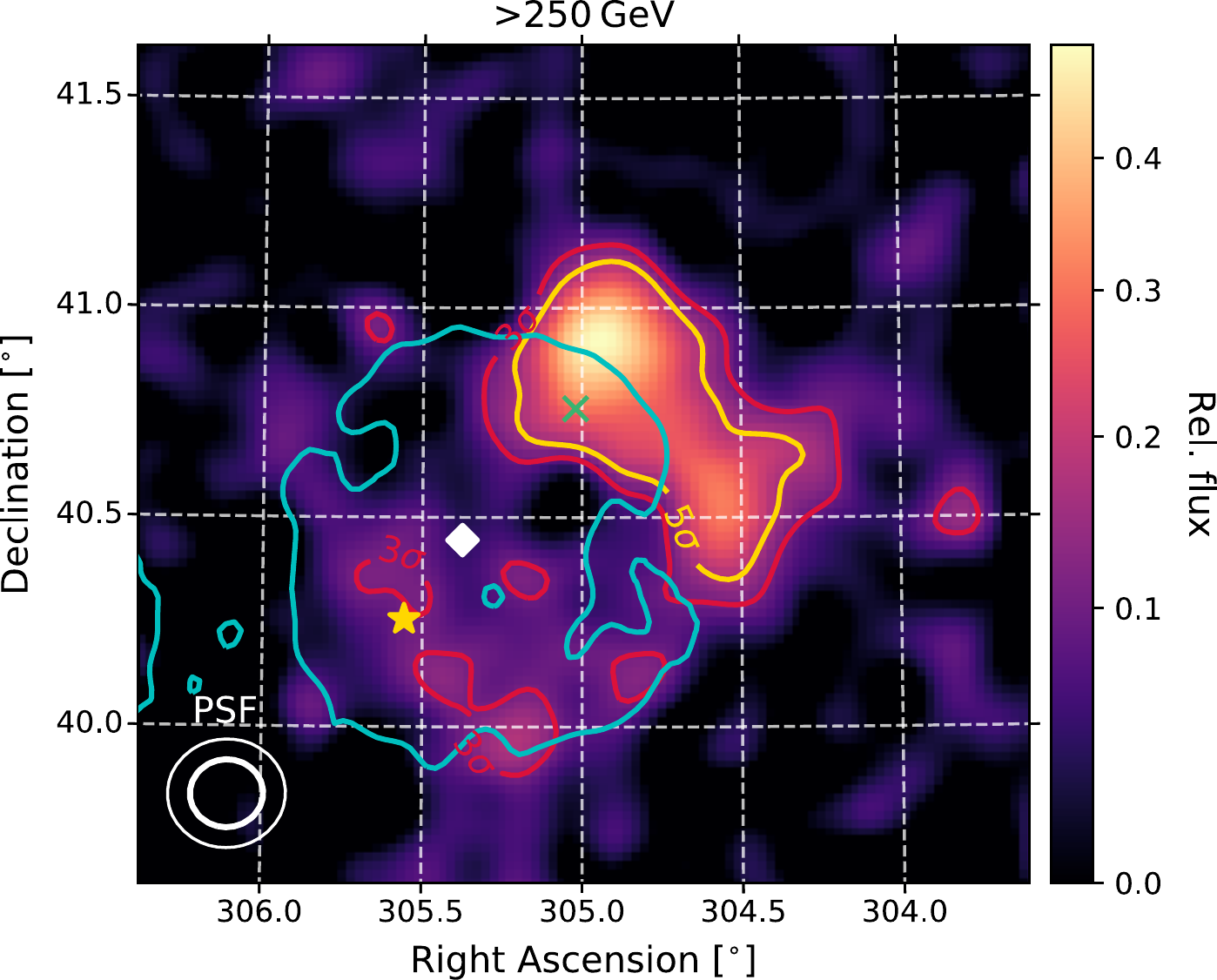}
	\caption{Sky map in units of relative flux (excess over background) of the $\gamma$-Cygni region as observed by MAGIC $E>\,250\,$GeV. Regions exceeding the 3$\sigma$ (5$\sigma$) local, pre-trial TS significance for a point source are indicated by red (yellow) contours. The cyan line is the 400\,K contour of the 408\,MHz observation by the CGPS. The white diamond marks the position of the PSR 2021+4026, the green X the position of VER\,J2019+407, and the yellow star the position of Sadr ($\gamma$-Cyg star; mag=2.2). The inlay in the lower left corner shows the 39\% and 68\% containment contours of the MAGIC PSF.} 
	\label{fig:MAGIC_overall}
\end{figure}

The image of the $\gamma$-Cygni SNR obtained with the MAGIC telescopes above 250\,GeV is displayed in Fig. \ref{fig:MAGIC_overall} in terms of the relative flux. Relative flux is defined as the residual or excess events divided by the background events. The event count map and the background map were smoothed with a Gaussian kernel of $\sigma=0\fdg 055$, corresponding to the size of the PSF. Additionally, the background map was scaled to have the same median pixel values of the event count map for the area outside of the defined exclusion region (Sect. \ref{sec:MAGIC_ana}). Figure \ref{fig:MAGIC_overall} also shows the 3 and 5\,$\sigma$ boundaries of the test statistic map, which for each pixel contains the deviation from a Gaussian distribution of the background only hypothesis based on 500 random representations of the background map. This $TS_{point}$ value gives a hint of the local significance for a point source in each pixel\footnote{Not to be confused with the significance estimates based on likelihood ratios obtained with SkyPrism and listed in Table \ref{tab::spec_results}}. The cyan lines are the 400\,K contours of the 408\,MHz observation by the CGPS \citep{taylor_canadian_2003}. In the following, we will continue using this surface brightness level since it agrees best with published extension values in the radio band and is the level right above the brightness of the surroundings.

The emission observed by MAGIC is extended and patchy. The most prominent feature in the map is an extended region along the north-western rim of the radio shell. It seemingly consists of a bright roundish component centred on the rim and an adjunct arc-like appendix extending beyond the radio shell towards the west. The map shows faint emission areas inside the southern part of the radio shell, which, however, are below the detection level for a point source.

Since the roundish emission in the north exhibits a higher surface brightness than the adjunct appendix, it seems implausible to account for both with just one source component. The former could be well modelled by a radially symmetric Gaussian with the position and the extension as free parameters. For the latter we used the sector of an annulus centred at the centre of the radio shell with the inner radius, the outer radius, the angular position of the centre (mathematically positive w.r.t. the decreasing RA axis), and the central angle as free parameters. However, it is difficult to determine the exact shape of the emission to the west of the SNR shell based on a visual inspection of the MAGIC skymap. Hence, we also applied a Gaussian to model this emission region. For the fit we assume a spectral, power-law index of $\Gamma=-2.8$ agreeing with the value reported by \citet{abeysekara_very_2018}.

Since the Gaussian source in the north of the SNR is better defined, we scanned its position and extension first. The scan resulted in a minimum at ($\alpha,\, \delta$; J\,2000)=$(304\fdg 89\pm0\fdg 01_{\rm stat},\, 40\fdg 84\pm0\fdg 01_{\rm stat})$ and an extension of $\sigma=0\fdg 16\pm0\fdg 01_{\rm stat}$. At this position we checked for a possible eccentricity, but the improvement over the radially symmetric Gaussian was not significant. The source was detected with a significance of 17.1\,$\sigma$ (in the absence of any other model component). In the following we will refer to this source as MAGIC\,J2019+408.

Including MAGIC\,J2019+408 in the source model, we scanned the parameters of the arc-like source. When modelled with the annulus segment, it is detected at a significance of 10.3$\sigma$ and the best-fit parameters are $0\fdg 45\pm0\fdg 03_{\rm stat}$ for the inner radius, $0\fdg 27\pm0\fdg 03_{\rm stat}$ for extension (outer - inner radius), $5\fdg 00\pm0\fdg 03_{\rm stat}$ for the positional angle, and $36\fdg 58\pm0\fdg 03_{\rm stat}$ for the central angle. 

When modelling the emission towards the west of the SNR with a second Gaussian instead of the arc model, the best-fit position was at ($\alpha,\, \delta$; J\,2000)=$(304\fdg 51\pm0\fdg 02_{\rm stat},\, 40\fdg 51\pm0\fdg 02_{\rm stat})$ with extension of $\sigma=0\fdg 12\pm0\fdg 01_{\rm stat}$. The detection significance was 9.2$\sigma$. We will address the question of model selection between these two alternatives in Sect. \ref{sec:Mod_Spec} after having examined the $\gamma$-ray emission over a wider energy range taking into account observations with \textit{Fermi}-LAT. A search for point-sources in addition to MAGIC\,J2019+408 and the arc or second Gaussian did not lead to any significant detection.

\subsection{Energy dependent morphology}
\label{sec:EnDepMorph}

Even though the observations with MAGIC provide a more precise image of the source at hundreds of GeV to TeV than previous observations, its morphology still differs from previously published {\it Fermi}-LAT skymaps. To study the evolution of the emission over the entire GeV to TeV range, we analysed the data set from the {\it Fermi}-LAT, which contained 1.5 times the observation time of the latest extended source catalogue \citep{ackermann_search_2017}. We split the {\it Fermi}-LAT and MAGIC data into two energy ranges each: 15\,--\,60\,GeV and 60\,--\,250\,GeV for the former and 250\,--\,500\,GeV and 500\,GeV\,--\,2.5\,TeV for the latter. 

\begin{figure*}
\centering
\includegraphics[width=17cm]{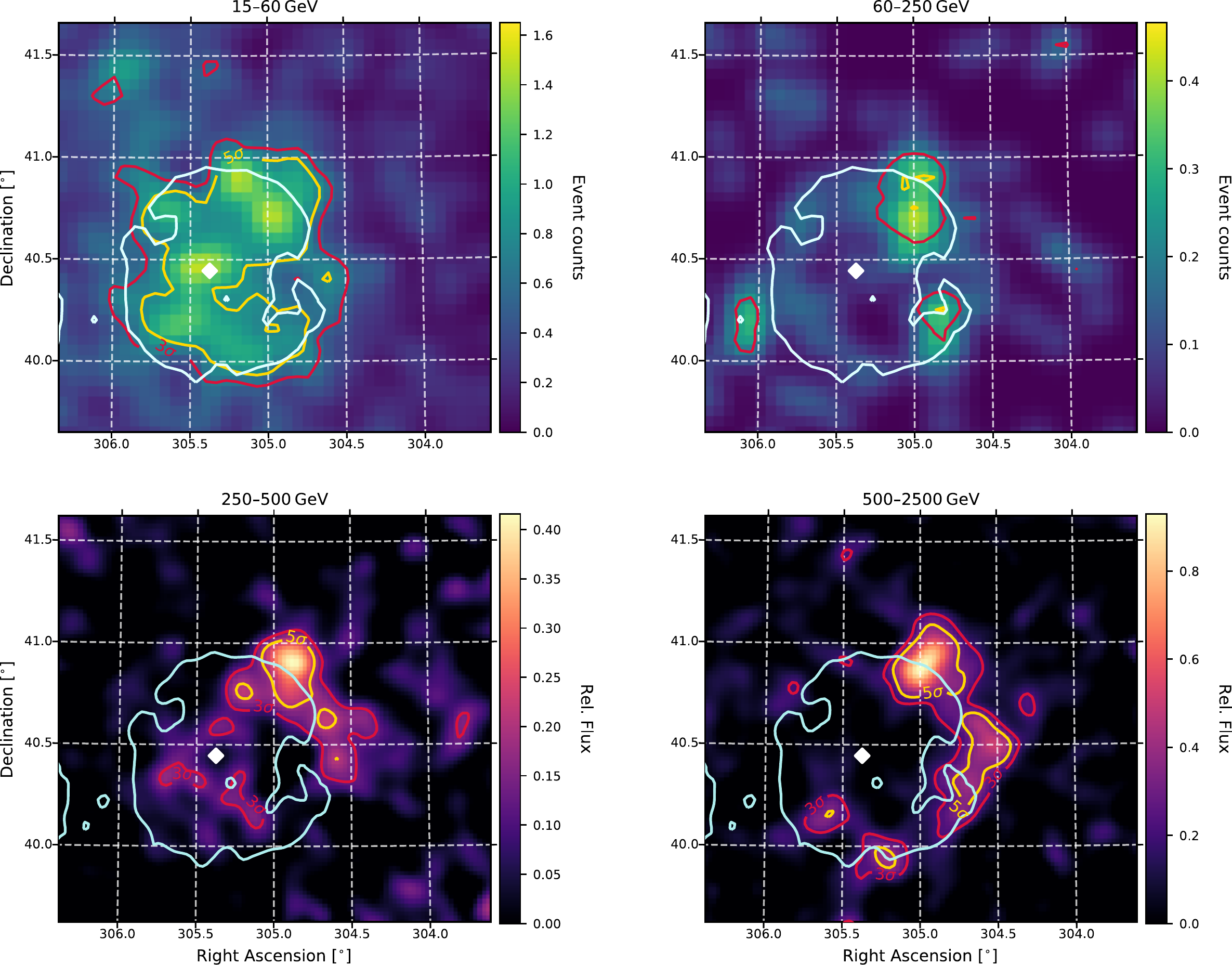}
\caption{Energy-dependent morphology of the $\gamma$-Cygni SNR. {\bf Upper left:} {\it Fermi}-LAT count map between 15 and 60\,GeV smoothed with a Gaussian kernel ($\sigma=0\fdg75$) and 3 and 5\,$\sigma$ contours of a point source search. The white line is the 400\,K contour of the 408\,MHz observation by the CGPS. The white diamond identifies the position of PSR 2021+4026. {\bf Upper right:} same as upper left but in the range from 60 and 250\,GeV. {\bf Lower left:} relative flux map observed by MAGIC at 250 to 500\,GeV together with 3 and 5\,$\sigma$ point source significance contours. The same radio contours as in upper panels are displayed in cyan. {\bf Lower right:} same as lower left but in the energy range from 500\,GeV to 2.5\,TeV.}
\label{fig:En_Dep_Morph}
\end{figure*}

The skymaps are shown in Fig. \ref{fig:En_Dep_Morph}. The {\it Fermi}-LAT maps display the event counts (smoothed with a Gaussian kernel with $\sigma$=0\fdg{}075) together with point source TS contours. The count images have the advantage of being free from any model dependency but do not allow the exclusion of sources by including them into the background model. Thus, we chose a lower limit of 15\,GeV to prevent the pulsar from dominating the image. TS contours are based on a point source search on top of the background source model described in Sect. \ref{sec:Fermi_ana}. The MAGIC skymaps are presented in terms of relative flux together with point source TS contours. In the 15 to 60\,GeV range the emission predominantly comes from a region agreeing with the SNR radio shell. The intensity is nearly uniform across the shell and the 3 and 5\,$\sigma$ local significance contours seemingly agree with the radio contours. On the contrary, at 60 to 250\,GeV the shell emission weakens and a slightly extended emission at the north-western rim stands out nearby MAGIC\,J2019+408. The position at which MAGIC observes the arc-like structure, however, does not show any significant emission.

MAGIC\,J2019+408 is the main component in the 250 to 500\,GeV map measured by MAGIC. Emission at the arc position is becoming visible but at a lower level compared to MAGIC\,J2019+408. The inside of the shell shows some faint emission. Finally, at 500\,GeV to 2.5\,TeV, the arc-like region brightens. The inside of the shell does not show any significant emission, instead, some emission towards the south of the shell becomes visible. Since this emission was not significant in the combined MAGIC data set and the TS contours are approximate Gaussian significances, it can only be considered a hint for emission.

\subsection{{\it Fermi}-LAT morphological results}
\label{sec:Fermi_results}

To quantify the {\it Fermi}-LAT results from Sect. \ref{sec:EnDepMorph}, the likelihood analysis presupposes a morphological template. To account for the extended uniform emission well visible in the 15 to 60\,GeV skymap, previous analyses of the {\it Fermi}-LAT data used a radially symmetric disk model. When fitting the {\it Fermi}-LAT data $E>5$\,GeV we used a radial symmetric disk with the position and extension being free parameters and obtained ($\alpha,\, \delta$; J\,2000)=$(305\fdg 24\pm0\fdg 02_{\rm stat},\, 40\fdg 49\pm0\fdg 02_{\rm stat})$ and a radius of $0\fdg 60\pm0\fdg 02$. This result is consistent with the disk model reported in \citet{ackermann_search_2017}. For the fit we used a spectral, power-law index of $\Gamma=2$, which is in the range of indices reported by the various {\it Fermi}-LAT catalogues.

\begin{figure}
	\includegraphics[width=\columnwidth]{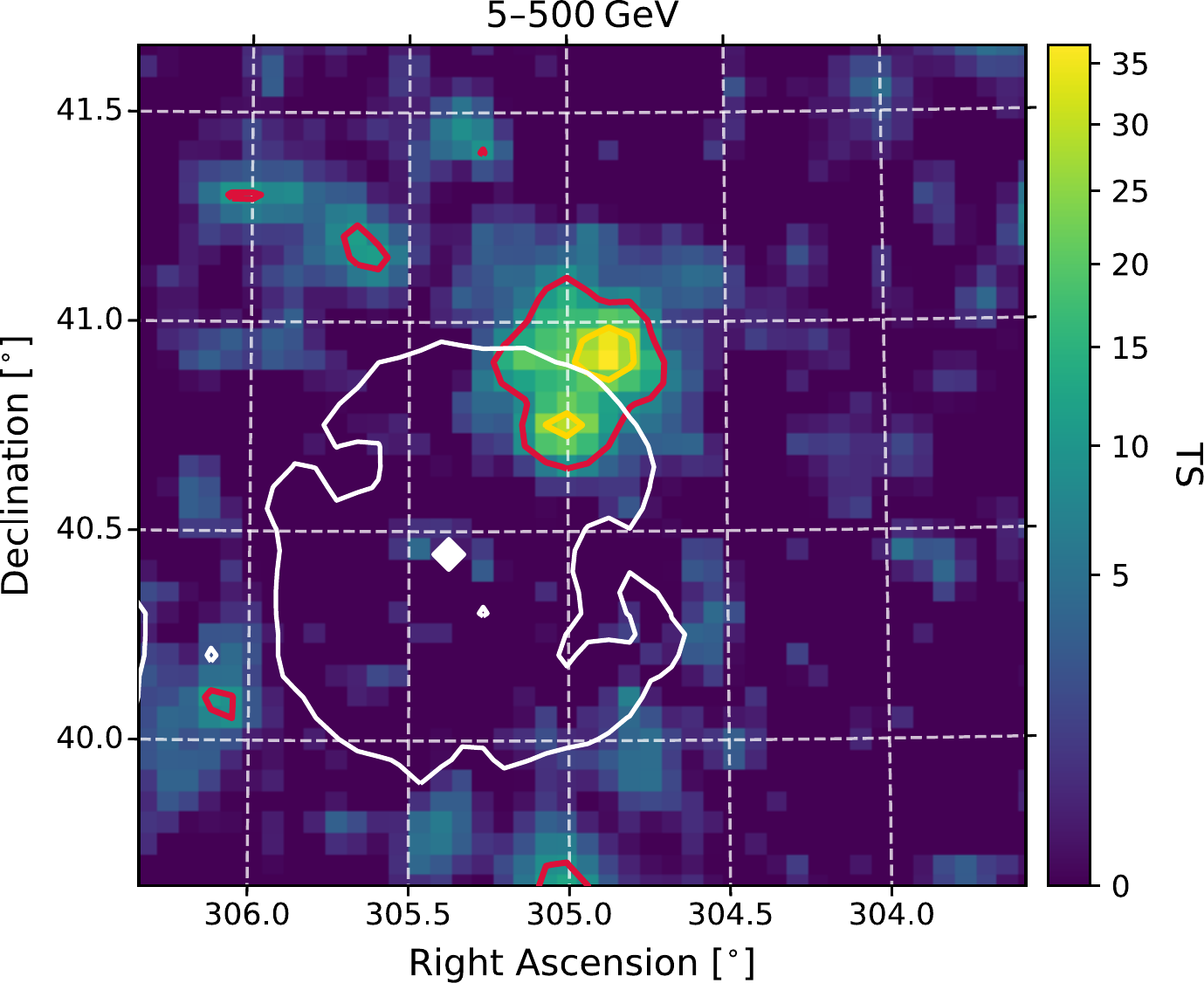}
	\caption{{\it Fermi}-LAT TS map for a point source search on top of the background sources specified in Sect. \ref{sec:Fermi_ana} and including the radio-based disk model. Regions exceeding the 3$\sigma$ (5$\sigma$) local, pre-trial TS significance for a point source are indicated by red (yellow) contours. The white line is the 400\,K contour of the 408\,MHz observation by the CGPS. The white diamond marks the position of PSR\,J2021+4026.} 
	\label{fig:Fermi_TSmap_HS}
\end{figure}

As mentioned in Sect. \ref{sec:sec_intro}, the estimated size of the disk varies between different {\it Fermi}-LAT studies (likely due to different energy ranges). All either partially or full encompass MAGIC J2019+408. However, since it is visible in the {\it Fermi}-LAT data ($E\,{>}\,60$\,GeV) and MAGIC data, it might be a distinct object from the disk. If it is not considered as such, photons might be misassigned to the disk pushing the fit towards a larger disk size. Assuming that the disk template is related to the SNR and given the approximate agreement between the TS contours and the radio contours in the 15\,GeV to 60\,GeV range, we consider it more plausible to base the disk model on the position and extension of the radio shell. The radially symmetric disk model is centred at ($\alpha,\, \delta$; J\,2000)=$(305\fdg 30,\, 40\fdg 43)$ with a radius of $0\fdg 53$. Fitting this model to the {\it Fermi}-LAT data resulted in remaining residual counts in the north-western region. Figure \ref{fig:Fermi_TSmap_HS} displays the TS map for a point source search on top of the radio disk. The contours suggest the presence of a source of similar morphology to MAGIC-J\,2019+408.

Assuming a point-source, the source search resulted in the detection at a 6.1$\sigma$ level ($TS=37$) at ($\alpha,\, \delta$; J\,2000)=$(304\fdg 92\pm0\fdg 02_{\rm stat}\pm0\fdg 02_{\rm sys},\, 40\fdg 87\pm0\fdg 02_{\rm stat}\pm0\fdg 02_{\rm sys})$. Using the Fermipy routines we additionally fitted the extension of the source simultaneously with the position using a radially symmetric Gaussian source model. The best fit is obtained for ($\alpha,\, \delta$; J\,2000)=$(304\fdg 98\pm0\fdg 03_{\rm stat}\pm0\fdg 02_{\rm sys},\, 40\fdg 87\pm0\fdg 03_{\rm stat}\pm0\fdg 02_{\rm sys})$ and an extension of $0\fdg 19\pm0\fdg 02_{\rm stat}\pm0\fdg 01_{\rm sys}$ (68\% containment; $\sigma=0\fdg13$). The extension of the source is significant with a TS value of 28.4. These values agree within errors with the position and extension of the MAGIC-J\,2019+408, so we associate the {\it Fermi}-LAT and MAGIC source. The fit was based on a spectral power-law index of $\Gamma=2$.

The two source, radio disk and MAGIC\,J2019+408, yield a log-likelihood value of $\ln\left(\mathcal{L}_{2\text{src}}\right)=-160597$, whereas the single extended disk model give $\ln\left(\mathcal{L}_{1\text{disk}}\right)=-160621$. However, both models are not nested and thus the LRT cannot be used to gauge the significance of the $\Delta\ln\left(\mathcal{L}\right)$. Instead we use the Akaike information criterion \citep[AIC,][]{Akaike1974} to assess the relative quality of the two model. The AIC estimate of a model is defined as

\begin{equation}
\label{Eq:AIC}
AIC = 2k - 2\ln\left(\mathcal{L}\right)\,,
\end{equation}

\noindent where $k$ is the number of degrees of freedom of the model and $\mathcal{L}$ the likelihood value. The model with the smaller AIC value is preferred. Since the radio disk is fixed the former model has 5 free parameter (norm of the disk and position [in 2-D], extension, and norm of the Gaussian) and the latter model has 4 (position, extension, and norm of the disk). Due to the $\Delta\text{loglike}=24$ improvement with only one additional parameter, the two-source model is preferred over the single larger disk.

\subsection{Common model and source spectra}
\label{sec:Mod_Spec}

\begin{figure*}
\sidecaption
	\includegraphics[width=12cm]{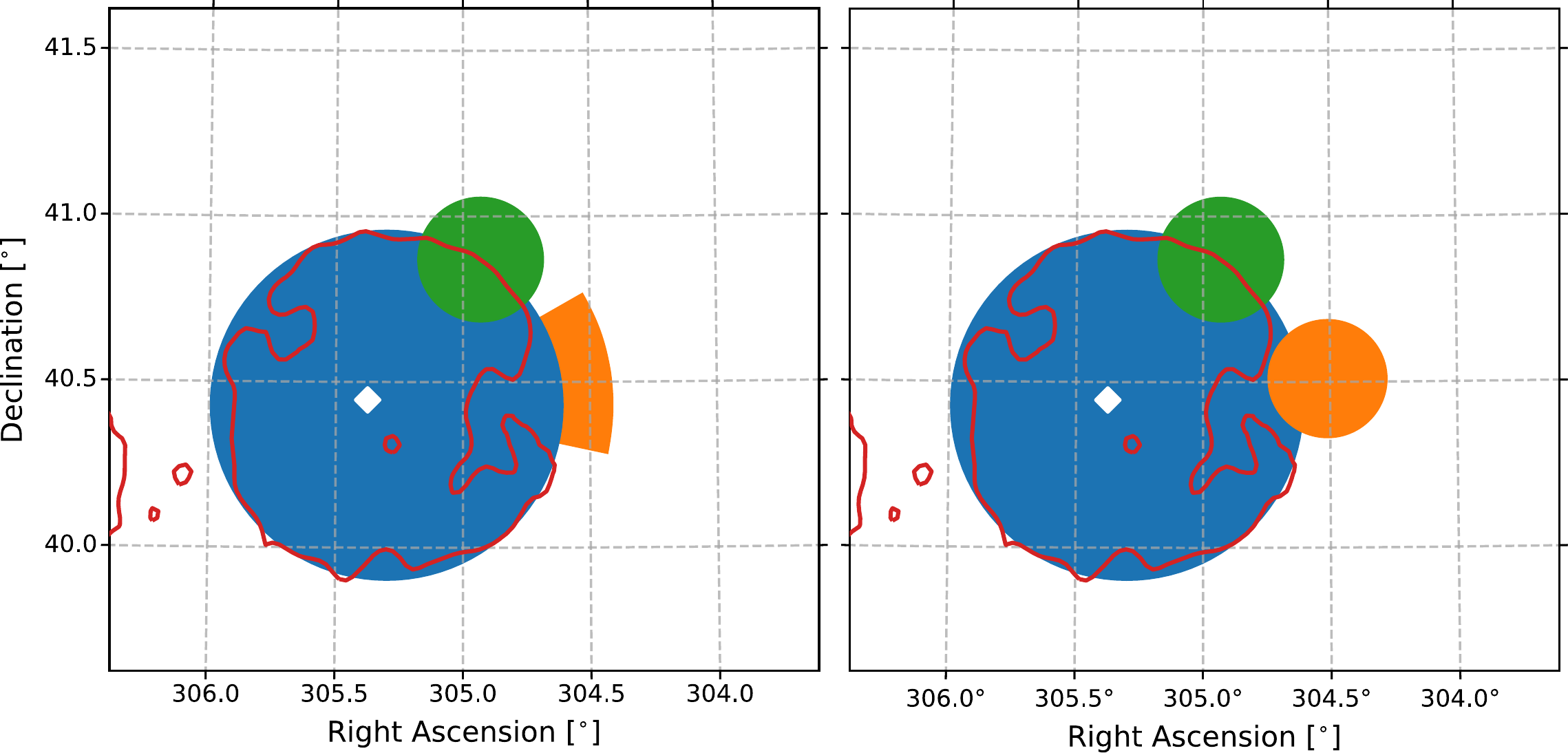}
	\caption{\textbf{Left:} Sketch of the spatial model used in the likelihood of the {\it Fermi}-LAT and MAGIC data analysis. The blue circle indicates the position and extension of the disk modelling the SNR shell, the green circle marks the position and 68\% containment radius of a Gaussian model of MAGIC\,J2019+408, and the orange annular sector was used for the arc region. The red radio contours and the position of PSR\,J2021+4026 (white diamond) are shown for reference. \textbf{Right:} Same as left figure, except with the second Gaussian as spatial model for the arc region.} 
	\label{fig:ModelPlot}
\end{figure*}

\begin{table*}
\caption{Spatial models used for the MAGIC and {\it Fermi}-LAT likelihood analysis.}             
\label{table:spatial_model}      
\centering                          
\linespread{1.15}\selectfont
\begin{tabular}{l c c c c c c}        
\hline\hline   
\multicolumn{1}{l}{Source name} & \multicolumn{1}{c}{Spatial model} & \multicolumn{2}{c}{Centred at} & \multicolumn{1}{c}{Extension} & \multicolumn{1}{c}{Positional angle} & \multicolumn{1}{c}{Central angle} \\    
	&  	& RA [deg] & Dec [deg] &  [deg]	& [deg]	& [deg]	\\    
\hline                        
SNR shell	& disk 	   & 305.30   & 40.43  & 0.53 (radius)	& - & - \\
MAGIC\,J2019+408	& Gaussian & 304.93   & 40.87  & 0.13 ($\sigma$) 	& - & - \\
Arc 		& annular sector & 305.30   & 40.43  & 0.15 ($r_{\text{out}}-r_{\text{shell}}$) & 7.0 & 33.0\\
Arc (alternative)	& Gaussian & 304.51   & 40.51  & 0.12 ($\sigma$) 	& - & - \\
\hline                                   
\end{tabular}
\linespread{1.0}\selectfont
\end{table*} 
To extract the spectra in the entire GeV to TeV range, we need to combine the findings in the Sect. above into a common source model. Particularly, based on the energy dependent brightness ratio between different parts around the SNR as described in Sect. \ref{sec:EnDepMorph}, in the following we consider the $\gamma$-ray emission in the region to consist of three components: the interior of the SNR shell, MAGIC\,J2019+408, and emission west of the shell (arc/second Gaussian). The position and parameters of the models for each source are given in Table \ref{table:spatial_model} and sketched in Fig. \ref{fig:ModelPlot}. 

For the emission from the SNR shell, we kept the disk based on the radio observation. This disk partially overlaps with MAGIC\,J2019+408 and the arc region, thus we updated the parameters of the MAGIC analysis including the disk. For MAGIC\,J2019+408 it resulted in ($\alpha,\, \delta$; J\,2000)=$(304\fdg 89\pm0\fdg 01_{\rm stat}\pm0\fdg 04_{\rm sys},\, 40\fdg 88\pm0\fdg 01_{\rm stat}\pm0\fdg 02_{\rm sys})$ and $\sigma=0\fdg 13\pm0\fdg 01_{\rm stat}\pm0\fdg 02_{\rm sys}$. To model MAGIC\,J2019+408 consistently for both instruments, we further averaged the MAGIC and {\it Fermi}-LAT location and extension of the Gaussian weighted with the inverse of their variances (statistical and systematic uncertainties added in quadrature): ($\alpha,\, \delta$; J\,2000)=$(304\fdg 93,\, 40\fdg 87)$ and $\sigma=0\fdg 13$. 

For the annular sector template representing the arc we fixed the inner radius at the radius of the shell (0\fdg53 from the SNR centre) and rescanned the other parameters. The best-fit values were $0\fdg 15\pm0\fdg 04_{\rm stat}\substack{+0.27 \\ -0.08}_{\rm sys}$ for extension, $7\fdg 00\substack{+4.0 \\ -2.6}_{\rm stat}\substack{+8.0 \\ -3.6}_{\rm sys}$ for the positional angle, and $33\fdg 0\pm6\fdg4_{\rm stat}\substack{+14.0 \\ -13.0}_{\rm sys}$ for the central angle. When supplying a second Gaussian as an alternative to the arc model. The best-fit was obtained for ($\alpha,\, \delta$; J\,2000)=$(304\fdg 49\pm0\fdg 02_{\rm stat}\pm0\fdg 05_{\rm sys},\, 40\fdg 56\pm0\fdg 02_{\rm stat}\pm0\fdg 03_{\rm sys})$ and $\sigma=0\fdg 12\pm0\fdg 02_{\rm stat}\pm0\fdg 04_{\rm sys}$. Fitting these sources to the {\it Fermi}-LAT and MAGIC data results in the detection significances stated in Table \ref{tab::spec_results}.

The model consisting of two Gaussians yields a slightly better log-likelihood value of $\ln\left(\mathcal{L}_{2\text{gauss}}\right)=-36303$ compared to the MAGIC\,J2019+408 and Arc model resulting in $\ln\left(\mathcal{L}_{\text{gauss}+\text{arc}}\right)=-36306$. Both models are not nested and we use the AIC (Eq. \eqref{Eq:AIC}) to access the relative model quality. Both models have the same number of free parameters: position (2D), extension, and the normalisation for the Gaussian and extension, positional angle, central angle, and flux normalisation for the Arc. The $\Delta~AIC$=-6 can be converted to a probability estimate via $\exp\left(\Delta~AIC/2\right)$ \citep[see e.g.][]{burnham_multimodel_2004}. Accordingly, the arc model is 6\% as likely as the second Gaussian, which is ${<}2\sigma$ in terms of Gaussian probability and thus the arc model cannot be rejected based on this statistical test. Due to the same number of free parameters in both models and the small difference in the log-likelihood, other information based model selection criteria (e.g. Bayesian information criterion) will give similar results. In the following, we will provide the spectra for both source models. However, in Sect. \ref{sec:Interpr_mod} we develop a theoretical model based on CRs escaping the SNR into a homogeneous medium. The annular sector is a better spatial representation for this model. Since both spatial templates are statistically equal, for the comparison against the theoretical model we will employ the annular sector. Nonetheless, it is important to note that the Gaussian model does not alter the understanding of emission outside the SNR shell. The centre of the Gaussian is clearly situated outside of the SNR radio shell with the centre of the Gaussian being $0 \fdg 1$ away from the edge of the SNR shell.

We extract the spectra of the source components from the MAGIC and {\it Fermi}-LAT data for the spatial components as defined in Table \ref{table:spatial_model}. For the {\it Fermi}-LAT data we used 9 bins per decade over the range from 5 to 500\,GeV and the MAGIC data were binned into 4 bins per decade ranging from 250\,GeV to 12.5\,TeV. To consider the effect of the energy migration, the Fermi science tools and the SkyPrism analysis adjust an assumed spectral model via a forward folding procedure. As spectral model we used a power law of the form

\begin{equation}
 \dv{N}{E}=N_{0}\left(\frac{E}{E_{0}}\right)^{-\Gamma}\,,
\end{equation}
with the photon index $\Gamma$, the normalization constant $N_{0}$, and the scaling energy (or pivot energy) $E_{0}$. The best fit results together with their uncertainties are summarised in Table \ref{tab::spec_results}. The choice of the spatial template for the arc region changes the spectra of the shell and MAGIC\,J2019+408 at a level below the statistical uncertainties. The contributions to the systematic uncertainties estimated as described in Sect. \ref{sec:Sys_Uncertainties} and listed in Table \ref{tab:sys_unc_spectra} are added in quadrature. Since, the spectral models are power-laws, the systematic uncertainty on the energy scale is converted to an uncertainty on the flux normalisation and added to it. We checked for a possible curvature of the spectrum of all sources by fitting a log-parabola spectrum (power-law exponent becomes $\Gamma+\beta\log(E/E_{0})$ with the curvature factor $\beta$). For none of the components and instruments a significant curvature was detected (improvement over power law based on a likelihood ratio test led to a significance $<3\sigma$).

\begin{table*}
	\caption{
     Results of the spectral analysis from the {\it Fermi}-LAT and MAGIC analysis for each source component. All sources were best fit with a power-law function with flux normalisation $N_{0}$, spectral index $\Gamma$, and scaling energy $E_{0}$. The model for fitting the {\it Fermi}-LAT data included the model for the arc region (either annular sector or Gaussian), but was found not to be significant.}      
   \label{tab::spec_results}
   \center
   \linespread{1.15}\selectfont
   \begin{tabular}{l c c c c}
      \hline\hline
      \multicolumn{1}{l}{Source name} 	& \multicolumn{4}{c}{MAGIC}\\    

      									& $N_{0}$ [$\mathrm{TeV}^{-1}\,\mathrm{cm}^{-2}\,\mathrm{s}^{-1}$] & $\Gamma$ & $E_{0}$ [TeV] & Det. Sign. [$\sigma$]\\
      \hline
      SNR Shell   						& $\left(10 \pm 2_{\mathrm{stat}}\substack{+6.7 \\ -3.5}_{\mathrm{sys}}\right)\,\times\,10^{-13}$ & $-2.55\pm0.16_{\mathrm{stat}}\substack{+0.30 \\ -0.25}_{\mathrm{sys}}$ & 1.0 & 6.1\\
      MAGIC\,J2019+408   				& $\left(10.0 \pm 0.9_{\mathrm{stat}}\substack{+6.0 \\ -3.5}_{\mathrm{sys}}\right)\,\times\,10^{-13}$ & $-2.81\pm0.10_{\mathrm{stat}}\substack{+0.21 \\ -0.19}_{\mathrm{sys}}$ & 1.0 & 16.7\\
      Arc (annular sector) 				& $\left(3.9 \pm 0.7_{\mathrm{stat}}\substack{+2.6 \\ -1.5}_{\mathrm{sys}}\right)\,\times\,10^{-13}$ & $-3.02\pm0.18_{\mathrm{stat}}\substack{+0.22 \\ -0.20}_{\mathrm{sys}}$ & 1.0 & 10.1\\
      Arc (Gaussian model)  			& $\left(5.2 \pm 0.8_{\mathrm{stat}}\substack{+3.6 \\ -2.2}_{\mathrm{sys}}\right)\,\times\,10^{-13}$ & $-2.99\pm0.16_{\mathrm{stat}}\substack{+0.22 \\ -0.22}_{\mathrm{sys}}$ & 1.0 & 10.3\\
      
      \hline
      									&	\multicolumn{4}{c}{{\it Fermi}-LAT}\\
      \hline      
      SNR Shell							& $\left(37 \pm 2_{\mathrm{stat}}\substack{+4.6 \\ -4.0}_{\mathrm{sys}}\right)\,\times\,10^{-10}$ & $-2.11 \pm 0.06_{\mathrm{stat}}\pm 0.01_{\mathrm{sys}}$ & 0.05 & 23.2\\
      MAGIC\,J2019+408					& $\left(9.8 \pm 1.8_{\mathrm{stat}}\substack{+1.1 \\ -1.0}_{\mathrm{sys}}\right)\,\times\,10^{-10}$ & $-1.86 \pm 0.13_{\mathrm{stat}} \pm 0.01_{\mathrm{sys}}$ & 0.05 & 8.9\\
      \hline
   \end{tabular}\\  
   \linespread{1.0}\selectfont
\end{table*}

To better compare the data against theoretical model predictions, we additionally compute data points (flux in narrow energy bands). For the {\it Fermi}-LAT data we rebin the 9 bins per decade to 2 bins between 5\,--\,13.9\,GeV, 2 bins between 13.9\,--\,64.6\,GeV, and 2 bins between 64.6\,--\,500\,GeV.  Thus, the energy bins are much wider ($\gtrsim4$\,times) than the energy resolution ($\Delta~E\,{<}\,5\%$) and, even though the data points are not deconvolved, the correlation between the points can be considered negligible. Whereas, for the MAGIC data, the energy resolution (15\,--\,20\%) is of the same order as the bin width, and the correlation between the data points needs to be taken into account. Hence, we used the forward folding technique for data points proposed in \citet{vovk_spatial_2018}, in which the differential flux points are interpreted as the breaks of a broken power-law with multiple energy breaks. The arc was not detected in the {\it Fermi}-LAT data and thus only spectral upper limits could be computed. The upper limits (ULs) of the {\it Fermi}-LAT data are 95\% confidence UL using the semi-Bayesian method of the science tools following \citet{helene_upper_1983} and the MAGIC UL are following the method by \citet{rolke_limits_2005}. In Sect. \ref{sec:gamma_ray_spectrum} these data points are compared against a model curve.

\subsection{Discussion of the observational results}
\label{sec:obs_discussion}

Regarding the disk, given the agreement of the position and extension of the emission detected by {\it Fermi}-LAT with the radio shell, a random coincidence seems implausible. As explained above, fitting the {\it Fermi}-LAT data resulted in a slight offset from the radio position and extension, but the fit of the {\it Fermi}-LAT data might be affected by the presence of MAGIC\,J2019+408. Hence, a disk model agreeing with the radio shell seems reasonable. Our spectral results are in agreement with previously published results on this source by the {\it Fermi}-LAT collaboration considering the difference in the area of the disk models.

MAGIC J\,2019+408 is present in the MAGIC and {\it Fermi}-LAT data. Its position is offset with respect to VER\,J2019+407 (0\fdg{}07 north of the latter) but still compatible ($\sim2\sigma$ discrepancy considering combined errors). \citet{aliu_discovery_2013} did not observe the arc structure, and not considering it as a distinct source may explain the different positions. The fact that the extension of VER\,J2019+407 is significantly larger and the VERITAS collaboration claimed an asymmetric source when updating their results in \citet{abeysekara_very_2018} supports this assumption. To highlight the different location and morphology, we gave this source an identifier different from the VERITAS one. In the {\it Fermi}-LAT energy range the spectrum of MAGIC J\,2019+408 is slightly harder than the one of the shell, consistent with the findings of \citet{fraija_gigaelectronvolt_2016}. The authors analysed {\it Fermi}-LAT data in a narrower energy range and performed a point-source search on top of the larger disk model from the 3FGL (whereas we used a physically-motivated radio-based model). Accordingly, they obtained a TS map different from ours (Fig. \ref{fig:Fermi_TSmap_HS}), associated the excess emission with VER J2019+407, and extracted the spectrum at the corresponding position. For MAGIC\,J2019+408 the possibility for an extended source unrelated to the SNR cannot be ruled out. However, X-ray and radio data do not show any hint for a possible Galactic counterpart such as a pulsar powering a wind. Additionally, the spectral agreement with the SNR interior, particularly in the energy range of {\it Fermi}-LAT, further supports the assumption of a connection with the SNR. 

The arc-like region is detected by MAGIC only, though the VERITAS skymaps in \citet{weinstein_cygnus_2015} and \citet{abeysekara_very_2018} show hints for an extended emission stretching out towards the west of the SNR. The differences in morphology can be understood as a result of the differences in the observation time ($t_{\text{MAGIC}}\,{\sim}\,2\times t_{\text{VERITAS}}$) and different methods for reconstructing the background emission (Exclusion region for MAGIC vs. ring background model for VERITAS), of which the ring background faces issues with extended sources \citep{berge_background_2007}. The exclusion region method is only insensitive to emission regions larger than twice the wobble distance ($1\fdg2$ for our MAGIC data). We can thus not rule out that the arc-like structure is the residual of a much larger complex such as the Cygnus Cocoon. However, given that the arc traces the rim of the SNR and its spectrum agrees with that of MAGIC\,J2019+408 at TeV energies, the association with the $\gamma$-Cygni SNR is very plausible. 

The PSF of the HAWC experiment does not allow resolving substructures in the $\gamma$-Cygni region making a comparison with the MAGIC results difficult. HAWC determined the centre of the emission around the SNR close to the centre of the shell \citep{abeysekara_2hwc_2017}, whereas from the MAGIC high energy skymap one would expect it to be shifted towards the north-west. This suggests additional emission surrounding the SNR likely towards the south of the shell, a region not well covered by the MAGIC observations due to the presence of Sadr. Still the steep spectral index measured by HAWC is in agreement with a softening of the spectrum between the energy range covered by {\it Fermi}-LAT and MAGIC. 
   \section{Interpretation and modelling}
\label{sec:Interpr_mod}

\subsection{Leptonic or hadronic emission}

The radio emission proves the presence of high-energy electrons inside the shell, which can also be the origin of the $\gamma$-ray emission via inverse Compton scattering (IC) or bremsstrahlung radiation. Due to the low plasma density of 0.2\,cm$^{-3}$ inside the SNR shell (Table \ref{table:snr_properties}), the former emission will dominate over the latter even when only considering a CMB photon field. The high energy spectrum from the shell of $\gamma$-Cygni up to a few hundreds of GeV has a slope of ${\sim}\,E^{-2}$. In contrast, the average radio spectral index $\alpha_{\rm R}$ of 0.48\,--\,0.75 \citep{zhang_multi-frequency_1997, gao_sino-german_2011, kothes_catalogue_2006, ladouceur_new_2008} implies an electron spectrum between $\text{d}N/\text{d}E\,{\propto}\,E^{-1.96}$ and ${\propto}\,E^{-2.5}$ and thereby a harder Inverse Compton (IC) spectrum in $\gamma$-rays. Hence, a leptonic scenario requires an additional breazk in the spectrum in the keV to GeV range to bring both observations into agreement. Such a break can naturally arise from electron cooling. To obtain a synchrotron cooling time shorter than the lifetime of the SNR, the magnetic field inside the SNR needs to be $B\,{\gtrsim}\,20$\,\textmu{}G.

If the emission outside the shell originates from IC as well, the morphology of the arc and MAGIC\,J2019+408 require either an enhancement of the radiation field in those regions or a specific guiding magnetic field creating an overdensity of electrons compared to other parts around the shell. Observations with the Infrared Astronomical Satellite \citep[IRAS; ][]{saken_iras_1992} indeed suggest a higher IR emission towards MAGIC\,J2019+408 at 25\,\textmu{}m and at 60\,\textmu{}m. However, the morphology of MAGIC\,J2019+408 does not agree with the IR structure and the peak of the former is offset with respect to the centre of the latter by ${\sim}\,0\fdg6$. Additionally, the parallaxes of identified IR sources in the vicinity of MAGIC\,J2019+408 \citep{collaboration_vizier_2018} suggest that at least part of the IR emission is farther away than the $\gamma$-Cygni system. The absence of non-thermal synchrotron radiation at MAGIC\,J2019+408 (skymaps in \citet{ladouceur_new_2008}) also renders the magnetic field scenario unlikely. Finally, the arc region is dark in both IR and synchrotron which speaks against an IC scenario as well.

Accordingly, the most likely leptonic scenario for the arc and MAGIC\,J2019+408 is bremsstrahlung emission. Like a hadronic scenario, it requires a local enhancement of the target gas density and is independent from the constraints above. Nevertheless, in this case the power-law index of the electron spectrum needs to change from $\Gamma\,{\sim}\,-3$ inside shell to $\Gamma\,{\sim}\,-2$ outside. Moreover, in order for bremsstrahlung to dominate over pion decay,  the accelerated electron-to-proton ratio has to be ${\gg}10^{-2}$, whereas the studies of multi-wavelength emission from several young SNRs point towards smaller ratios of $\approx10^{-3}$ or less \cite[see, e.g.,][]{Volk+2005,Morlino-Caprioli2012}. Additionally, theoretical predictions based on particle-in-cell simulations of collisionless shocks hint at values of electron-to-proton ratios of ${\lesssim}10^{-2}$ \citep{Park+2015}. 

In conclusion, even if a leptonic explanation cannot be completely ruled out, its realisation requires extreme conditions. Alternatively, the $\gamma$-ray emission can be explained in a hadronic scenario, which is not subject to the aforementioned constraints. Hence, in the following we develop a hadronic model to explain the data and accordingly assume that the bulk of emission is due to hadronic interactions.

\subsection{ Escaping or precursor?}  \label{sec:escaping}

If the emission from the arc region is indeed connected to the SNR, the emission beyond the SNR shell can be either due to the CR precursor in front of the shock or produced by particles escaping from the shock. 
The former interpretation seems unlikely for two different reasons. Firstly, the spectrum from the arc region is softer (at most similar given the uncertainties) than the one detected from the SNR interior. Using the linear theory with a spatially constant diffusion coefficient in the precursor, the spectrum upstream of the shock is given by:

\begin{equation} 
\label{eq:f_up}
 f_{\rm up}(x,p) = f_{\rm sh}(p) \, e^{-{u_{\rm sh} x/D_1}} \,,
\end{equation}
where $x$ is the distance upstream from the shock, $p$ the particle's momentum, $u_{\rm sh}$ is the shock speed, and $D_{1}$ the diffusion coefficient upstream \citep[see e.g.][]{Blasi2013}. If the spectrum at the shock is $f_{\rm sh} \propto p^{-\alpha}$ and the diffusion coefficient is $D_1(p) \propto p^{\beta}$ (in general $\beta>0$ and $\beta=1$ for Bohm diffusion), the spatially-integrated spectrum upstream is

\begin{equation}
\int_0^{\infty} f_{\rm up}(x,p) \dd x \propto p^{-\alpha+ \beta}\,.
\end{equation}
Hence, the spectrum from the arc region should be harder than the one inside the remnant unless the diffusion coefficient is constant in momentum, which would be difficult to explain from both observational and theoretical grounds.

The second argument comes from the comparison of the SNR age with the acceleration time. If the arc represents the shock precursor, the thickness of the arc $\Delta_{\rm arc}$ corresponds to the diffusion length $\lambda_p$ of particles with momentum $p$ upstream. Hence, we can estimate the diffusion coefficient at the central energy observed imposing $\lambda_p=\Delta_{\rm arc}\,{\simeq}\,D_1(p)/u_{\rm sh}$. At a distance of 1.7\,kpc, the extension of the arc is $\Delta_{\rm arc}\,{\sim}\,5$\,pc for central energy of all MAGIC data of ${\sim}\,800$\,GeV, corresponding to parent protons of ${\sim}8$\,TeV, leading to a diffusion coefficient upstream of the shock equal to

\begin{equation} \label{eq:D20TeV}
 D_1(8\, {\rm TeV}) \simeq \lambda_p u_{\rm sh} = 
       1 \times 10^{27} \left( \frac{\lambda_p}{4.5 \,{\rm pc}} \right)  
       \left( \frac{u_{\rm sh}}{10^3 \,{\rm km \,s^{-1}}} \right) \, {\rm cm^2 \,s^{-1}} \,.
\end{equation}

Using the test particle approach and following e.g. \citet{drury_introduction_1983}, from the diffusion coefficient we can estimate the acceleration time needed to produce particles at 8\,TeV as

\begin{equation} \label{eq:t_acc}
 t_{\rm acc}(p) = \frac{3}{u_1-u_2} \left( \frac{D_1}{u_1} + \frac{D_2}{u_2} \right)
                        \simeq 8 \, \frac{D_1}{u_{1}^2}  \simeq 4  \times 10^{4} \, {\rm yr} \,,
\end{equation}

\noindent which is ${\sim}\,5$ times the estimated SNR age ($D_{2}$ is the diffusion coefficient downstream, $u_{2}$ the velocity of the downstream plasma, and $u_{1}$ the shock speed). 

A major uncertainty regarding this interpretation results from the unknown 3-D orientation of the $\gamma$-ray emission. The $\gamma$-ray data do not allow estimating the distance along the line-of-sight. Hence, our analysis may e.g. misassign emission belonging to the arc-region and situated outside of the SNR shell to our disk model if, in the 2-D projection, it is mapped onto the SNR shell. Consequently, a possible misalignment can conceal spectral differences between the arc-region and the shell. Accordingly, the precursor could not be excluded based on the spectral similarity. In that case however, the extension of the arc would be underestimated by our model, strengthening the argument of the acceleration time. In summary, a precursor scenario for the arc region seems improbable and instead it is a region where particles escaping from $\gamma$-Cygni interact with the ISM.

Using the observed extension of the arc we can put a lower limit to the external diffusion coefficient, $D_{\rm out}$, assuming that particles located in the arc started escaping at the beginning of the ST phase. Considering the typical energy of 8\,TeV for CR protons and reasonable values for the SNR parameters (see Table \ref{table:snr_properties} and $M_{\rm ej}=5 M_\odot$), we estimate the diffusion coefficient via the length ($\lambda_{\rm diff} = (6D_{\rm out} t)^{1/2}$; factor 6 assumes 3 dimensions):

\begin{equation} \label{eq:Dout_approx}
 D_{\rm out}(8\, {\rm TeV}) = \frac{\lambda_{\rm diff}^2}{6t} \simeq \frac{(R_{\rm SNR} + \Delta_{\rm arc} - R_{\rm ST})^2}{6\left(t_{\rm SNR} - t_{\rm ST}\right)}
     \simeq 2 \times 10^{27} \rm cm^2 \, s^{-1} \,,
\end{equation}

\noindent where $R_{\rm ST} = ( M_{\rm ej}/\rho_0)^{1/3}$ and $t_{\rm ST} = E_{\rm SN}^{-1/2} M_{\rm ej}^{5/6} \rho_{0}^{-1/3}$, while  $\Delta_{\rm arc} = 0\fdg15 \simeq 5$\,pc.
The estimated $D_{\rm out}$  is ${\sim}\,\num{7e2}$ times smaller than the average Galactic diffusion coefficient at 8\,TeV as obtained from direct CR measurements, i.e. $D_{\rm Gal} \approx 6 \, \beta \,10^{28} (E/{\rm GeV})^{1/3} {\rm cm^2 \, s^{-1}}$ \citep[see e.g.][]{trotta_constraints_2011, yuan_propagation_2017}. Here and in the rest of the paper we assume a slope of 1/3 typical for Kolmogorov turbulence. If the turbulence were determined by other processes, the slope could be different. However, a different slope would not affect our main conclusions. Even when considering the large systematic uncertainty on the extension of the arc ($\Delta_{\rm arc}$ would instead be 12\,pc), $D_{\rm out}$ is still ${\sim}\,\num{3e2}$ times smaller than the average $D_{\rm Gal}$. Note that changing the SNR parameters in the range reported in Table~\ref{table:snr_properties} (always assuming that the estimated extension matches the real one), the ratio $D_{\rm Gal}/D_{\rm out}$ ranges between ${\sim}$\numrange{8e1}{3e3}. Note, that the considerations in this section would equally apply to a leptonic scenario with the only difference that the parameters would be tested at a parent particle energy of ${\sim}\,2$\,TeV instead of 8\,TeV. Furthermore, the results would also hold when taking the off-set between the SNR shell and the centre of the alternative Gaussian model for the arc-region for $\Delta_{\rm arc}\sim0 \fdg 1$.

The result obtained in Eq. (\ref{eq:Dout_approx}) represents an underestimation for two reasons: 1) we assume that the arc mainly extends orthogonally to the line of sight. If this is not true, $\Delta_{\rm arc}$ would be larger than 5 pc, resulting in a larger $D_{\rm out}$. 2) the arc could represent an over-dense region. If beyond such region the density drops to a lower value, $\gamma$-rays could be undetectable even if CR have diffused beyond $\Delta_{\rm arc}$. A more thorough discussion on this point will be given in section \ref{sec:discussion}.

In the escape scenario, CRs are expected to escape radially symmetrically or, in case of a dominant main magnetic field, to escape mainly along the magnetic field direction. Emission beyond the shell should thus not be solely seen in the direction of the arc. A straightforward explanation could be that the arc has a larger density than the rest of the circumstellar medium (see discussion in Sect. \ref{sec:gamma_ray_spectrum}). Additionally, a large scale magnetic field oriented in the direction of the arc may cause a larger density of particles escaping into the arc region and can explain why emission is concentrated there \citep{nava_anisotropic_2013}. Indeed, the radio emission indicates that the magnetic field is directed along the arc. In fact, the radio shell is not homogeneous but presents two main lobes in the south-east and north-west, the latter roughly agreeing with the direction of the arc. Because the shock acceleration theory predicts a larger efficiency for the parallel shock configuration (i.e. $\vec{B} \parallel \vec{u}_{\rm sh}$) \citep{berezhko_emission_2002, caprioli_acceleration_214}, the two radio lobes could be interpreted as polar caps. In such a situation another bright region can be expected on the opposite side of the SNR with respect to the arc, but again a low target gas density could impede its visibility. 

Furthermore, the non-detection of emission on the opposite side of the SNR may partly result from the telescope pointing position chosen to avoid the influence of the bright star Sadr. The opposite site of the SNR is about ${\gtrsim}\,1\degr$ away from our pointing positions, where the acceptance of the MAGIC telescopes decreases to ${\lesssim}\,1/2$ of the full sensitivity. Nonetheless, MAGIC should have detected the emission if it had the same surface brightness within the covered energy range as the arc region, hence the emission in the south might be weaker, distributed over wider area, or have a different spectral energy distribution. The hints for emission at the south of the SNR shell visible in figure \ref{fig:En_Dep_Morph}, and that HAWC determined the centre of the SNR to agree with the centre of the shell, still could be a sign for the existence of a counterpart to the arc around the southern shell.    

\subsection{A simplified approach for particle propagation}    \label{sec:model}
In this section we model the propagation of accelerated particles inside and outside the SNR in order to properly calculate the $\gamma$-ray emission. We follow the derivation proposed by \cite{Celli+2019b} (the reader is referred to that paper for further details). For simplicity we assume spherical symmetry inside and outside the remnant. The transport equation for accelerated protons in spherical coordinates is 

\begin{equation} \label{eq:transport}
 \frac{\partial{f}}{\partial{t}} + u \frac{\partial{f}}{\partial{r}} = 
 \frac{1}{r^2} \frac{\partial}{\partial r} \left[ r^2 D \frac{\partial{f}}{\partial{r}}\right]
 + \frac{1}{r^2} \frac{\partial (r^2 u)}{\partial r}  \frac{p}{3} \frac{\partial f}{\partial p} \,,
\end{equation}

\noindent where $u$ is the advection velocity of the plasma and $D$ the diffusion coefficient. 
The former is obtained from the SNR evolution. Because $\gamma$-Cygni is clearly in the ST phase, we describe its evolution using the ST solution in the case of expansion inside uniform medium with density $\rho_0$. The shock position $R_{\rm sh}$ and the shock speed as a function of time are

\begin{eqnarray} \label{eq:ST}
 R_{\rm sh}(t) = \left( \xi_0 \frac{E_{\rm SN}}{\rho_0} \right)^{1/5} t^{2/5} \,,	\\
 \label{eq:ST2}
 u_{\rm sh}(t) = \frac{2}{5} \left( \xi_0 \frac{E_{\rm SN}}{\rho_0} \right)^{1/5} t^{-3/5} \,,
\end{eqnarray}

\noindent where $\xi_0= 2.026$ (for a monatomic gas with specific heat ratio $\gamma=5/3$).
The internal structure of the SNR is determined by adopting the linear velocity approximation \citep{Ostriker-McKee1988}, in which the gas velocity profile for $r < R_{\rm sh}$ is given by 

\begin{equation} \label{eq:u(r,t)}
 u(r,t) =  \left(1-\frac{1}{\sigma} \right) \frac{u_{\rm sh}(t)}{R_{\rm sh}(t)} r \,,
\end{equation}

\noindent where $\sigma$ is the compression factor at the shock. The radial profile of the gas density in the SNR interior (needed to calculate the $\gamma$-ray emission) is also given by the ST solution and can be well approximated by the following polynomial with respect to the self-similar variable  $\bar{r} = r/R_{\rm sh}(t)$ \citep{sedov_similarity_1959},

\begin{equation}  \label{eq:n_in}
 \rho_{\rm in}(\bar{r}) =  \sigma \, \rho_0 \left( a_1 \bar{r}^{\alpha_1} + a_2 \bar{r}^{\alpha_2} + a_3 \bar{r}^{\alpha_3} \right)  \,,
\end{equation} 

\noindent where $\rho_0$ the upstream density and the parameters $a_k$ and $\alpha_k$ are obtained from a fitting procedure which gives: $a_1= 0.353$, $a_2= 0.204$, $a_3=0.443$, $\alpha_1= 4.536$, $\alpha_2= 24.18$, and $\alpha_3= 12.29$ \citep{Celli+2019b}.

In the subsequent sections we will solve the transport equation~(\ref{eq:transport}) using two different approximations, one for particles confined inside the remnant and one for the escaping particles.

\subsection{CR distribution at the shock}    \label{sec:f_shock}
Following \cite{ptuskin_spectrum_2005}, we assume that the shock converts the bulk kinetic energy to relativistic particles with an efficiency $\xi_{\rm CR}$, which is constant in time. The distribution function of CRs accelerated at the shock is determined by the DSA and is predicted to be a power law in momentum up to a maximum value, $p_{\max,0}$. In a simplified form we can write the spectrum at the shock as

\begin{equation} \label{eq:f_0}
 f_0(p,t) = \frac{3 \, \xi_{\rm CR} u_{\rm sh}(t)^2 \rho_0}{4 \pi \, c (m_p c)^4  \Lambda(p_{\max,0}(t))} 
 		\left( \frac{p}{m_p c} \right)^{-\alpha}  \Theta \left[ p_{\max,0}(t)-p \right] \,,
\end{equation}

\noindent where $m_p$ is the proton mass, $\Theta$ is the Heaviside function while $\Lambda(p_{\max,0})$ is the function required to normalize the spectrum such that the CR pressure at the shock is $P_{\rm CR} = \xi_{\rm CR} \, \rho_0 u_{\rm sh}^2$. We keep the power law index $\alpha$ as a free parameter in order to fit the $\gamma$-ray data. We notice, however that DSA predicts $\alpha$ to be equal or very close to 4. 

The maximum momentum at the shock is a function of time and its calculation requires a correct description of the evolution of the magnetic turbulence. However, this is a non-trivial task because the magnetic turbulence depends on the self-generation by the same particles as well as by damping effects and wave cascades. A comprehensive description of all these effects does not exist yet. Hence, here we use the general assumption, often used in the literature \cite[see, e.g.][]{Gabici+2009}, that the maximum momentum increases linearly during the free expansion phase and decreases as a power law during the ST phase:

\begin{equation} \label{eq:pmax0}
 p_{\max,0}(t) =
  \begin{cases} 
   p_M \left( t/t_{\rm ST} \right)     & \text{if } t < t_{\rm ST} \\
   p_M \left( t/t_{\rm ST} \right)^{-\delta}     & \text{if } t > t_{\rm ST} \,,
  \end{cases}
\end{equation}

\noindent where $p_M$, the absolute maximum momentum reached at $t=t_{\rm ST}$, and $\delta >0$ are treated as free parameters of the model. 

Inverting equation(\ref{eq:pmax0}) we can also define the {\it escaping time}, when particles with momentum $p$ cannot be confined any more and start escaping from the remnant, 

\begin{equation} \label{eq:tesc}
 t_{\rm esc}(p) = t_{\rm ST} \left( p/p_M \right)^{-1/\delta} \,.
\end{equation}

It is also useful to define the {\it escaping radius}, namely the radius of the forward shock when particles with momentum $p$ start escaping, i.e.

\begin{equation} \label{eq:Resc}
 R_{\rm esc}(p) = R_{\rm sh}\left(t_{\rm esc}(p) \right) \,.
\end{equation}

\noindent The particle distribution evolves in a different way before and after $t_{\rm esc}(p)$, as we discuss below.

\subsection{Distribution of confined particles}    \label{sec:f_conf}
When $t < t_{\rm esc}(p)$, particles with momentum $p$ are confined inside the SNR and do not escape. A reasonable approximation for the distribution of these confined particles, that we call $f_c$ from here on, can be obtained from equation(\ref{eq:transport}) neglecting the diffusion term. This approximation is accurate for $p \ll p_{\max,0}(t)$, but we will show in a moment that in the test-particle case the diffusion inside the SNR does not play an important role. The simplified transport equation is

\begin{equation} \label{eq:transport_in}
 \frac{\partial{f_{\rm c}}}{\partial{t}} + u \frac{\partial{f_{\rm c}}}{\partial{r}} = 
 \frac{1}{r^2} \frac{\partial (r^2 u)}{\partial r}  \frac{p}{3} \frac{\partial f_{\rm c}}{\partial p} 
\end{equation}

\noindent and the solution can be easily obtained using the method of characteristics, accounting for the plasma speed inside the SNR as approximated by equation (\ref{eq:u(r,t)}). The solution can be written as follows \cite[see also][]{ptuskin_spectrum_2005}

\begin{equation} \label{eq:f_c}
 f_{\rm c}(t,r,p) = f_0 \left( \left( \frac{R_{\rm sh}(t)}{R_{\rm sh}(t')} \right)^{1-\frac{1}{\sigma}} p, t'(t,r) \right) \,,
\end{equation}

\noindent where $t'(t,r)$ is the time when the plasma layer located at the position $r$ at time $t$ has been shocked, namely 

\begin{equation} \label{eq:t_sh}
 t'(t,r) = \left(\rho_0 \, \xi_0^{-1} E_{SN}^{-1}\right)^2 r^{10} t^{-3}.
\end{equation}

We simplify equation(\ref{eq:f_c}) using equations (\ref{eq:f_0}) and (\ref{eq:ST}) and neglecting the mild dependence of $\Lambda(p_{\max})$ on $t$, and get

\begin{equation} \label{eq:f_c2}
 f_{\rm c}(t,r,p) = f_0(p,t) \left( \frac{t'}{t} \right)^{2\alpha (\sigma -1)/5\sigma -6/5}
 	      \Theta\left[ p_{\max}(t,r) - p \right] \,.
\end{equation}

\noindent The function $p_{\max}(t,r)$ is the maximum momentum of particles at position $r$ and time $t$ and it is equal to the maximum momentum of particles accelerated at time $t'$ diminished by adiabatic losses:

\begin{equation} \label{eq:pmax}
 p_{\max}(t,r) = p_{\max,0}(t') \left( \frac{R_{\rm sh}(t')}{R_{\rm sh}(t)} \right)^{1-\frac{1}{\sigma}} 
 	             =  p_{\max,0}(t) \left( \frac{t'}{t} \right)^{\frac{2}{5} \frac{\sigma-1}{\sigma} - \delta}  \,,
\end{equation}

\noindent where the last step uses equation (\ref{eq:pmax0}).
Interestingly, assuming test-particle DSA, where $\alpha= 3\sigma/(\sigma-1)$, the distribution function of confined particles has only a mild dependence on $r$ through the normalisation factor $\Lambda(p_{\max})$. In such a case neglecting diffusion in first approximation is justified because $\partial_r f_{\rm c} \approx 0$.

\subsection{Distribution of escaping particles}    \label{sec:f_esc}
When $t > t_{\rm esc}(p)$, particles with momentum $p$ cannot be confined any more and start escaping. In previous works, the escape is assumed to be instantaneous, meaning that particles with momentum $p$ are immediately located outside the remnant at $t > t_{\rm esc}(p)$. While this assumption can be a valid approximation for studying the final CR spectrum released into the Galaxy, it is invalid in the case of $\gamma$-Cygni as we aim for describing the early phase of the escape process in a region close to the SNR.

An approximate solution for time $t > t_{\rm esc}(p)$ can be obtained assuming that particles are completely decoupled from the SNR evolution and only diffuse. The evolution is therefore described by equation(\ref{eq:transport}) dropping all terms including $u$:

\begin{equation} \label{eq:transport_out}
 \frac{\partial{f_{\rm esc}}}{\partial{t}} = 
 \frac{1}{r^2} \frac{\partial}{\partial r} \left[ r^2 D \frac{\partial{f_{\rm esc}}}{\partial{r}}\right] \,.
\end{equation}

\begin{figure}
\begin{center}
\includegraphics[width=0.48\textwidth]{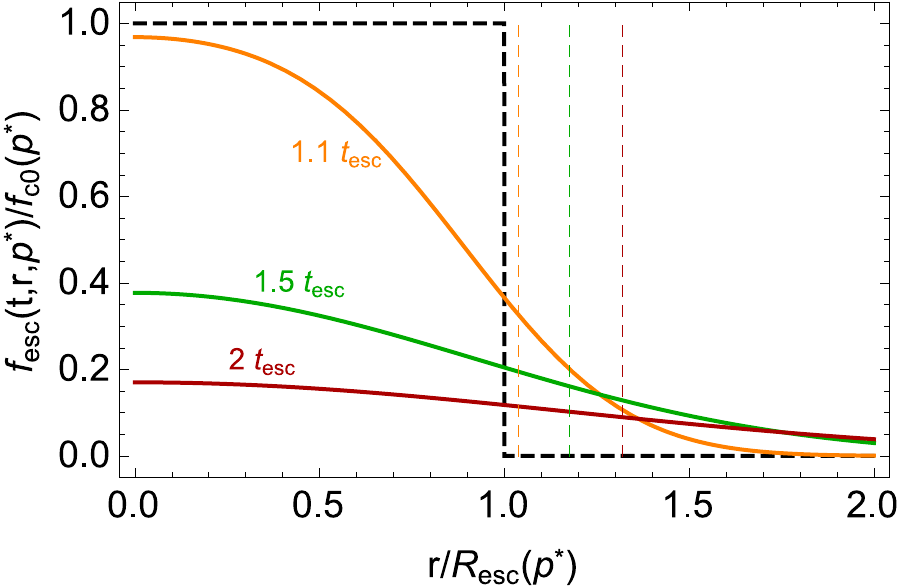}
\end{center}
\caption{Distribution of escaping particles at one arbitrary fixed momentum, $p^*= 10$ TeV, as a function of the radial coordinate normalised to $R_{\rm esc}(p^*)= 13$ pc. Different lines refer to different times in unit of $t_{\rm esc}(p^*)= 4000$ yr, as labelled, and the vertical dashed lines correspond to the shock position at those times. We assume $D_{\rm out}=D_{\rm Gal}/100$, $\delta=2.2$ and $p_M= 100$ TeV.}
\label{fig:f_esc1}
\end{figure}

\begin{figure}
\begin{center}
\includegraphics[width=0.48\textwidth]{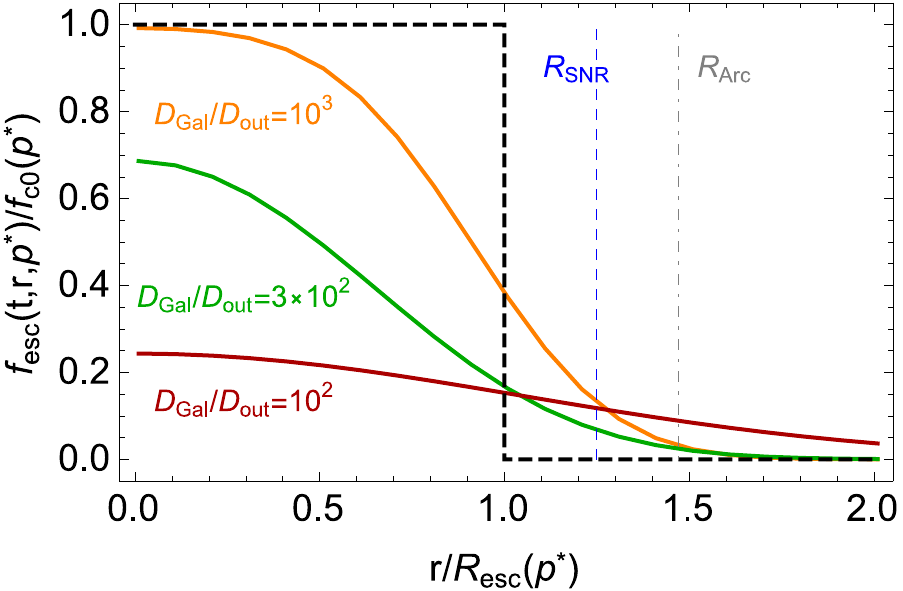}
\end{center}
\caption{Distribution of escaping particles at $p^*= 10$ TeV/$c$ and $t=t_{\rm SNR}$ as a function of the radial coordinate normalised to $R_{\rm esc}(p^*)= 13$ pc. Different lines refer to different values of the diffusion coefficient, as labelled. The vertical  lines correspond to the shock position (dashed) and to the arc external edge (dot-dashed) as observed now. The remaining parameters are the same as in figure \ref{fig:f_esc1}.}
\label{fig:f_esc2}
\end{figure}

This equation needs to be solved with the initial condition: $f_{\rm esc}(t_{\rm esc}(p), r,p) = f_{\rm c}(t_{\rm esc}(p), r, p) \equiv f_{c0}(r,p)$ for $r < R_{\rm esc}(p)$ and 0 elsewhere. The diffusion coefficient in the medium outside the SNR, $D_{\rm out}$, is assumed to be spatially constant and is an unknown of the problem that we want to constrain from observations. Inside the SNR the diffusion coefficient, $D_{\rm in}$, is in general different from the one outside, nevertheless, for simplicity we assume $D_{\rm in}= D_{\rm out}$ (and spatially constant). Such an approximation allows for an analytic solution of equation~(\ref{eq:transport_out}) via the Laplace transform \cite[see][for the full derivation]{Celli+2019b}. The final result is

\begin{eqnarray} \label{eq:fesc}
  f_{\rm esc}(t,r,p) 
    = \frac{f_{c0}(p)}{2}  \, \Theta\left[ t - t_{\rm esc}(p) \right] \times \hspace{2cm} \nonumber \\
    	\left\{ \frac{R_d}{\sqrt{\pi} \, r} 
  		\left( e^{-R_{+}^2} - e^{-R_{-}^2} \right)  		
    		+ {\rm Erf} \left( R_{+} \right) + {\rm Erf} \left( R_{-}\right)
	\right\}  \,,
\end{eqnarray}

\noindent where $R_{\pm} = \left(R_{\rm esc}(p) \pm r \right)/R_d(p)$, $R_d= \sqrt{4 D(p) \left(t -t_{\rm esc}(p) \right)}$, and Erf is the error function.
Examples of $f_{\rm esc}$  are plotted in Figures~\ref{fig:f_esc1} and \ref{fig:f_esc2} for different times and different values of the diffusion coefficient. For all plots we assume a strong shock ($\sigma=4$) and the test particle limit ($\alpha = 4$). 

When $D_{\rm in} \neq D_{\rm out}$ the leaking of particles from the remnant changes but the profile of the distribution function outside of the remnant remains essentially the same, being determined mainly by $D_{\rm out}$.

   \section{\texorpdfstring{$\gamma$}{gamma}-ray spectra}

\subsection{Emission from the SNR interior and arc} \label{sec:gamma_ray_spectrum}

Once the particle distribution is known at any position inside and outside the SNR, the calculation of $\gamma$-ray emissivity due to hadronic collisions is straightforward. The rate of emitted photons from a given region is

\begin{equation} \label{eq:F_gamma}
 \Phi_{\gamma}(E_{\gamma},t) 
 	\equiv \frac{\dd N_\gamma}{\dd E_\gamma\, \dd t} 
	= 4 \pi \int \frac{\dd \sigma(E_p,E_\gamma)}{\dd E_{\gamma}} J_p(E_p, t) \dd E_p \,,
\end{equation}

We parametrise the differential cross section following \cite{Kafexhiu+2014}. 
$J_p(E_p,t)$ is the spatially-integrated proton flux as a function of the kinetic energy $E_p$ and observation time $t$. It is connected to the momentum distribution function as $J_p(E_p) \dd E_p = \beta c F_p(p) \dd^3 p$, where $F_p$ is the proton distribution in momentum convoluted with the target density in the region of interest. In particular we distinguish two regions, the interior of the SNR and the fraction of the external spherical shell which include the arc. The corresponding distributions are

\begin{eqnarray} \label{eq:Fp}
 F_{p,\rm SNR}(p, t_{\rm SNR}) = \int_{0}^{R_{\rm SNR}} 4 \pi r^2 \, n_{\rm in}(r) f_{\rm in}(t_{\rm SNR},r,p) \, \dd r \\
 F_{p,\rm arc}(p, t_{\rm SNR}) = \eta_{\rm arc} \, n_{\rm arc} \int_{R_{\rm SNR}}^{R_{\rm SNR} + \Delta_{\rm arc}} 4 \pi r^2 \, f_{\rm esc}(t_{\rm SNR},r,p) \, \dd r \,,
\end{eqnarray}

\noindent where the particle distribution inside the SNR is $f_{\rm in} = f_c$ for $p < p_{\max,0}(t_{\rm SNR})$, while $f_{\rm in} = f_{\rm esc}$ otherwise, with $f_c$ and $f_{\rm esc}$ given by Eqs. (\ref{eq:f_c2}) and (\ref{eq:fesc}), respectively.
The gas distribution inside the remnant is given by Eq. (\ref{eq:n_in}), while in the arc we assume a constant density defined by $n_{\rm arc}$. The additional factor $\eta_{\rm arc}$ accounts for the spherical shell fraction which includes the arc region.

\begin{figure}
\begin{center}
\includegraphics[width=0.48\textwidth]{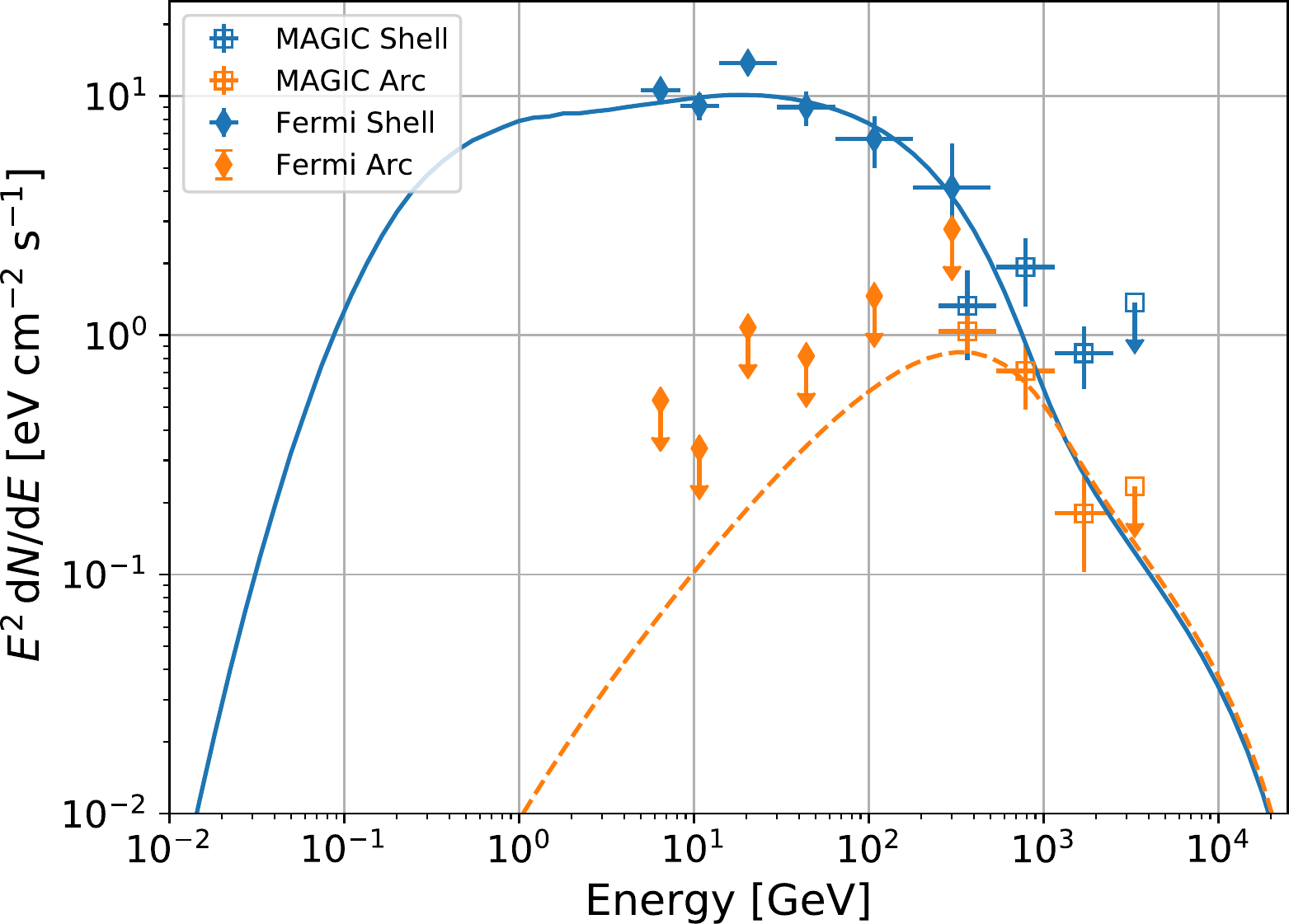}
\end{center}
\caption{Computed $\gamma$-ray flux due to $\pi^0$ decay from the SNR interior (blue solid line) and from the arc region (orange dashed line) based on our model compared against data points from our {\it Fermi}-LAT (diamonds) and MAGIC (squares) analysis.}
\label{fig:gamma_flux}
\end{figure}

Figure~\ref{fig:gamma_flux} shows our best fit to the observed $\gamma$-ray flux for both the SNR interior and the arc region. The parameters used to produce these curves are summarised in Table~\ref{table:2}. Parameters related to the SNR evolution are fixed to the values from Table~\ref{table:snr_properties}. All the other parameters are allowed to vary. The corresponding numbers in square brackets show the range of values resulting in curves, which are still in reasonable agreement with the data. 

At a first glance, the large number of free parameters (six if we exclude the ones related to the SNR) may suggest a strong degeneracy between them. Nevertheless, we can fix all the values with a reasonable small level of uncertainty because every parameter is connected to a specific feature of the spectrum.

First, the {\it Fermi}-LAT data from the remnant interior and the radio data fix the slope of the accelerated spectrum below TeV energies to be $\alpha \simeq 4.0$. Secondly, the normalisation of the $\gamma$-ray flux fixes the acceleration efficiency to be $\xi_{\rm CR} \simeq 4\%$.
In the energy range between 100 and 300 GeV the slope abruptly changes from ${\sim}E^{-2}$ to ${\sim}\,E^{-2.5}$. In our model this turning point defines the maximum energy of particles presently accelerated to be $E_{\max}(t_{\rm SNR})\,{\simeq}$\,1\,--\,3\,TeV. Noticeably, such energy is independently constrained by the {\it Fermi}-LAT upper limits on the flux from the arc region: to be compatible with the MAGIC data, the $\gamma$-ray flux from this region needs to have a maximum in the range 100\,--\,300 GeV. Such a maximum corresponds to $\gamma$-rays produced by the lowest energetic particles in the arc, which is very close to the maximum energy of particles accelerated now. 
Additional information is derived from the shape of the MAGIC spectrum that simultaneously determines $E_{\rm MAX}$ (the maximum energy reached at the beginning of the ST phase), $D_{\rm out}$ and $\delta$.
Our model predicts that the shape of the $\gamma$-ray emission from the SNR and from the arc should be very similar in the MAGIC band. Considering the uncertainties in the data, this is compatible with observations.

\begin{figure}[t]
\begin{center}
\includegraphics[width=0.48\textwidth]{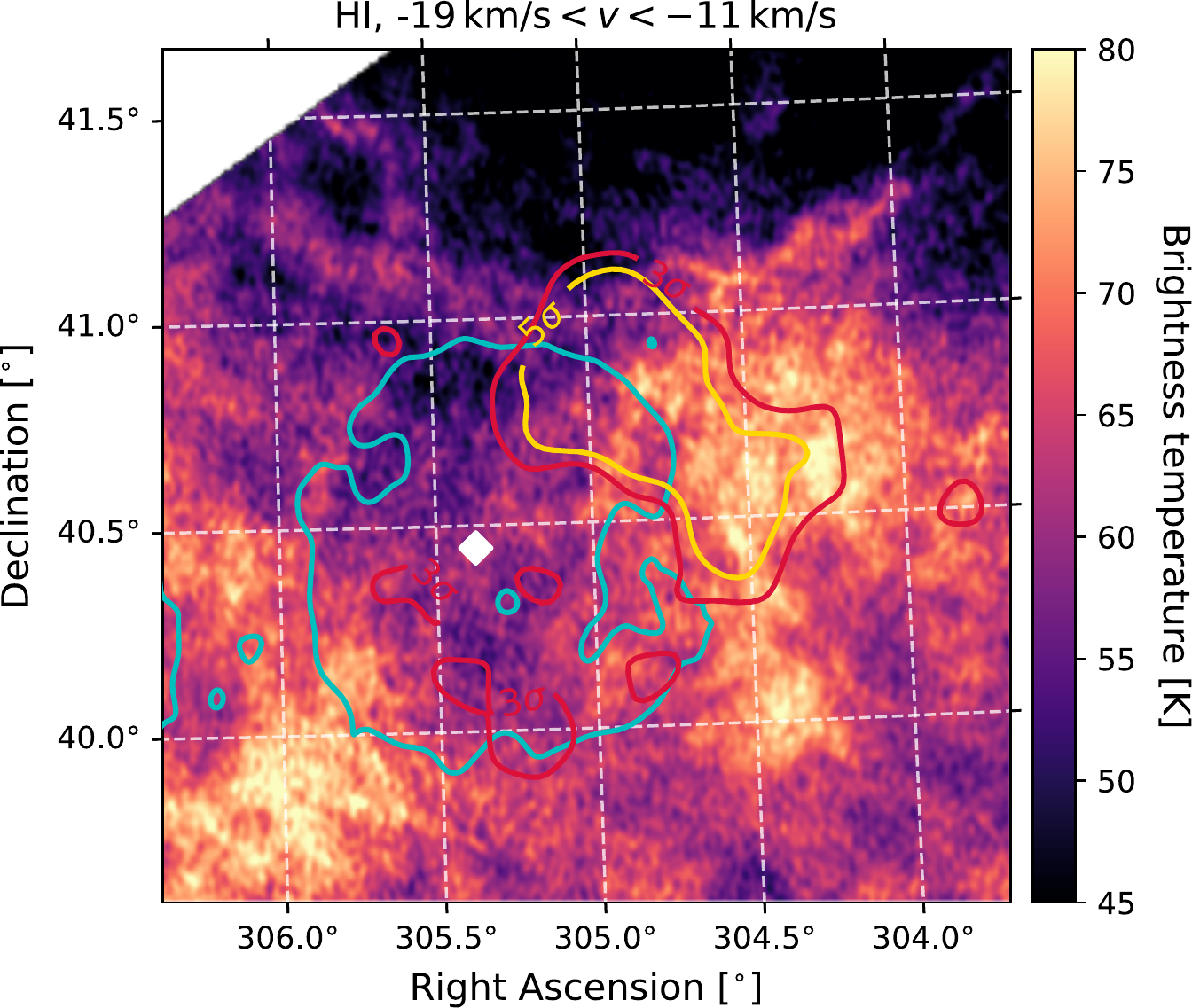}
\end{center}
\caption{HI emission at velocities between \SIrange{-19}{-11}{\kilo\metre\per\second} measured by the CGPS. The red and yellow lines show the 3 and 5 \,$\sigma$ significance contours of the $\gamma$-ray emission observed by MAGIC (Fig. \ref{fig:MAGIC_overall}). The Cyan lines are the \SI{400}{\kelvin} contours of the \SI{408}{\mega\hertz} continuum emission from the CGPS. The white diamond marks the position of PSR\,J2021+4026}
\label{fig:HI}
\end{figure}

Finally, the normalisation of the MAGIC data points of the arc spectrum sets the product of $\eta_{\rm arc} \times n_{\rm arc}$. Because the observed geometry suggests a filling factor $\eta_{\rm arc}\,{\sim}20\%$ (with some uncertainties due to a possible line-of-sight effect, see Sect. \ref{sec:escaping}) we also have an estimate of the target density in the arc region which has to be 1\,--\,2\,cm$^{-3}$. The targets inside the arc regions could be either the possible cavity wall reported by \citet{ladouceur_new_2008} or smaller clumps like the one found inside the north-west shell by \citet{uchiyama_asca_2002}. Indeed the location of the cavity wall surrounding the SNR shell to the north claimed by \citet{ladouceur_new_2008} at velocities between -19\,km/s and -11\,km/s coincides with the arc region (see Fig. \ref{fig:HI}). Even in the optically thin case the column density in this range reaches \SI{1e21}{\per\square\centi\metre} by far exceeding the column density required by our model by one order of magnitude at least.

We also note that some level of uncertainty is introduced by the parametrisation of the differential cross section used in Eq. (\ref{eq:F_gamma}). We tried all the four models considered in \cite{Kafexhiu+2014} (Geant 4.10, Pythia 8.18, SIBYLL 2.1 and QGSJET-1) but, for sake of clarity, we only show the results obtained with Pythia because for $E_{\gamma}\,{>}\,1$ GeV it gives a $\gamma$-ray flux roughly in between the maximum and minimum predictions (obtained with SIBYLL and Geant, respectively). At lower energies all the methods give essentially the same result. The uncertainty in the cross section mainly corresponds to a factor ${\sim}\,2$ difference in the target density of the arc or, equivalently, in the acceleration efficiency. This uncertainty is accounted for in the uncertainty interval shown with square brackets in Table~\ref{table:2}.

\begin{table*}
\caption{Value of parameters used to fit the $\gamma$-Cygni spectrum shown in Fig.~\ref{fig:gamma_flux}. The first block refers to the SNR properties, the second to the acceleration and escaping properties and the last one to the properties of the external medium (density times filling factor and diffusion coefficient). The number in square brackets shows the possible range of values that can still give reasonable fits to the data.}
\label{table:2}      
\centering                          
\begin{tabular}{c c c c c | c c c c | c c}        
\hline\hline                 
$E_{\rm SN}$   &   $M_{\rm ej}$   &   $t_{\rm SNR}$   &   $d$           &   $n_{0}$   &   
$\xi_{\rm CR}$  &   $\alpha$         &   $E_{\rm MAX}$  &   $\delta$    &    $\eta_{\rm arc} n_{\rm arc}$   &    $D_{\rm Gal}/D_{\rm out}$     \\    
\hline                        
 $10^{51}$ erg  &  $ 5\,{\rm M}_{\odot}$ & 7 kyr & 1.7 kpc & 0.2 cm$^{-3}$  & 3.8\%  &  4.0 &  78\,TeV  &  2.55  & 0.31 cm$^{-3}$  & 16    \\
 \multicolumn{5}{c|}{[see Table \ref{table:snr_properties}]}  & [3\%\,--\,7\%] &  [3.9\,--\,4.2] &  [20\,--\,250] &  [2.2\,--\,3.8] &  [0.25\,--\,0.45]   &  [10\,--\,35] \\
\hline                                   
\end{tabular}
\end{table*}

\subsection{Emission from MAGIC\,J2019+408} \label{sec:hotspot}

MAGIC\,J2019+408 has been detected both by {\it Fermi}-LAT and MAGIC. Remarkably the high energy spectrum is very similar to the one detected from the arc region, wheras in the {\it Fermi}-LAT band the slope is slightly harder than the emission from the overall SNR but still compatible with the latter. This finding suggests that at least part of the emission from the hot-spot should come from a region located inside the SNR, where low energy particles are also present. 

\begin{figure}[t]
\begin{center}
\includegraphics[width=0.48\textwidth]{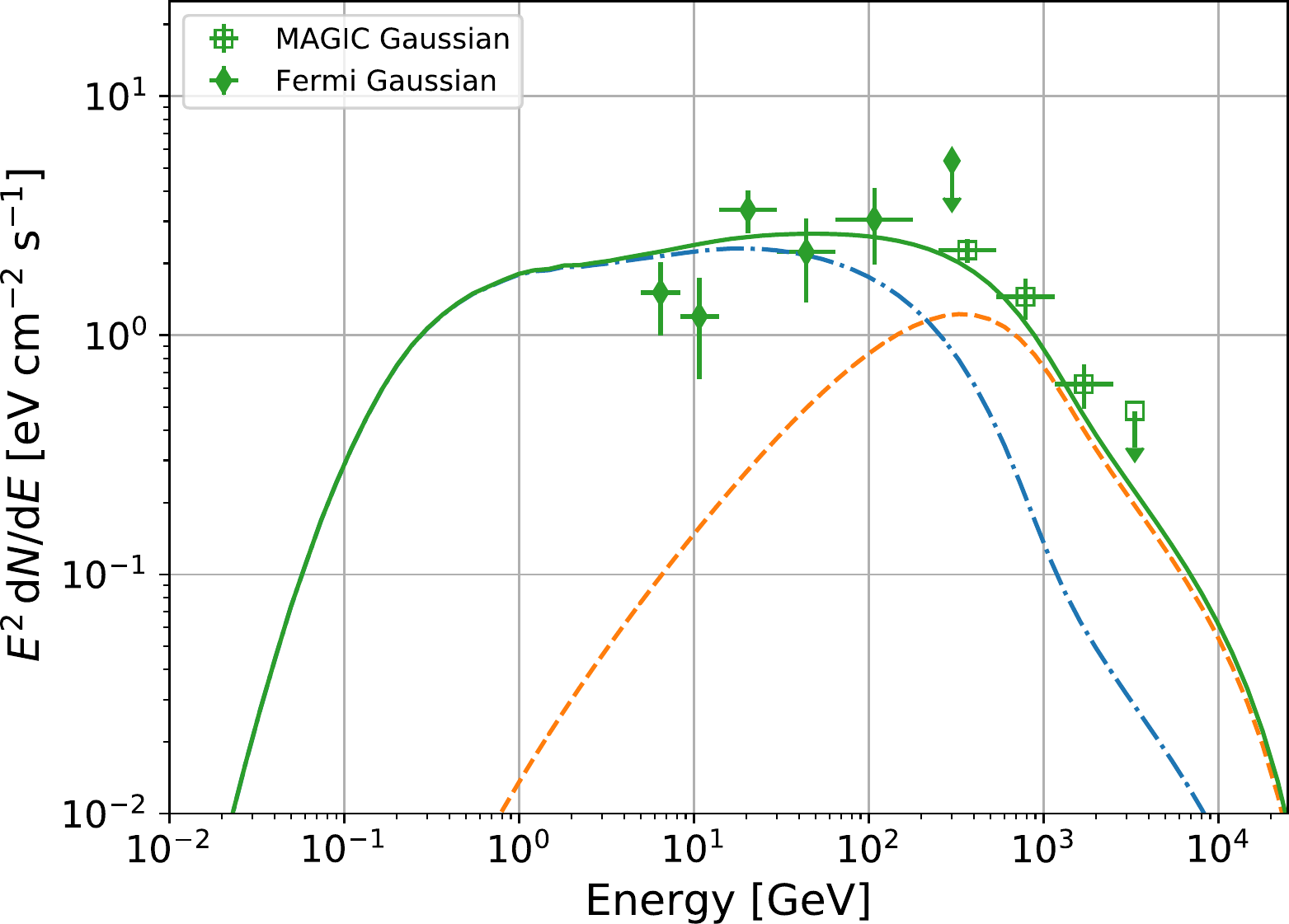}
\end{center}
\caption{Computed $\gamma$-ray flux due to $\pi^0$ decay from MAGIC\,J2019+408 based on our model compared against data points from our {\it Fermi}-LAT (diamonds) and the MAGIC (squares) analysis. The spectrum (solid, green line) is modelled as the sum of emission from the remnant interior (blue dash-dotted line) and exterior (orange dashed line).}
\label{fig:gamma_flux-HS}
\end{figure}

In the framework of the hadronic model proposed here, MAGIC\,J2019+408 can be understood as a combination of emission from the remnant interior plus a contribution from escaping particles located outside the SNR. In Fig.~\ref{fig:gamma_flux-HS} we show a possible fit to the data by combining these two components and only adjusting the normalisation via the target density. Remarkably this fit does not require any further tuning of the remaining parameters of the model and we keep the values reported in Table~\ref{table:2}. Considering that the angular extension of the hot-spot is $\sim 0.1^{\circ}$ and assuming a spherical geometry, the shown fit requires a density of $\sim45 $ cm$^{-3}$ both for the internal and the external contributions. Hence, MAGIC\,2019+408 could be due to an over-dense cloud partially engulfed by the SNR shock and partially still outside of it. Alternatively, MAGIC\,J2019+408 could result from two (or more) clouds, spatially separated, one inside and one outside the SNR, but located along the same line of sight.
It is worth mentioning that the estimated density of $\sim45 $ cm$^{-3}$ is close to the value of ${\sim}20 $ cm$^{-3}$ reported by \citet{mavromatakis_deep_2003} but in the East of the SNR. 

\begin{figure*}[t]
\begin{center}
\includegraphics[width=17cm]{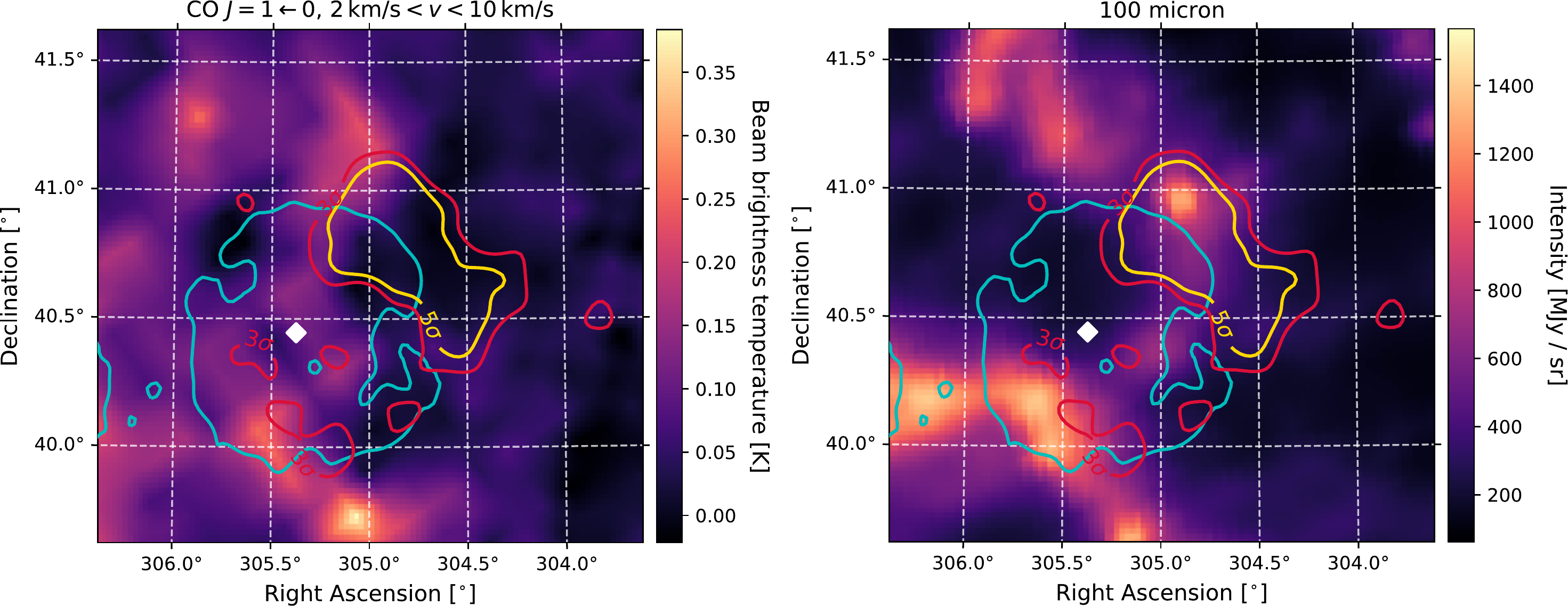}
\end{center}
\caption{
\textbf{Left:} Average of CO J${=}1{\leftarrow}0$ emission between 2\,km/s and 10\,km/s based on data from \citet{leung_complete_1992} interpolated on the coordinate grid of the MAGIC data. The cyan contours are the 400\,K contour of the 408\,MHz observation by the CGPS and the red and yellow contours are the 3\,$\sigma$ and 5\,$\sigma$ significance contours of the emission seen by MAGIC. The white diamond marks the position of the PSR\,J2021+4026.
\textbf{Right:} 100\,\textmu~m intensity map based on data from \citet{schlegel_maps_1998} interpolated on the coordinate grid of the MAGIC data. Contours and marker are the same as in the left panel.
}
\label{fig:CO_dust}
\end{figure*}

Even though no direct interaction between the SNR and a cloud has been reported for the north of the SNR, CO emission is present at a velocity range of 2 to 10\,km/s as shown by \citet{ladouceur_new_2008} based on CO J${=}1{\leftarrow}0$ observations by \citet{leung_complete_1992}. The left panel of Fig. \ref{fig:CO_dust} shows the SNR radio contours and MAGIC significance contours on top this map. The right panel shows how the same contours compare against infrared emission at 100\,\textmu{}m indicating the location of dust. The CO emission borders the MAGIC contours and the dust emission is even located inside of them. However, both, CO and dust, rather trace the dense core of interstellar clouds. The density of our model for MAGIC\,2019+408 of $\sim45 $ cm$^{-3}$ is small compared e.g. to the critical density of CO (${\sim}\,\SI{1.1e3}{\per\cubic\centi\metre}$) and due a typical dust-to-gas ratio of about $1:100$ little emission from dust is expected. Hence the absence of a direct morphological counterpart of MAGIC\,2019+408. Nonetheless, the $\gamma$-ray emission may interact with the outer layers of a potential cloud. 
If the target material of MAGIC\,2019+408 consists of HI entirely, the column density at the centre of a spherical cloud would be ${\sim}$\SI{5e20}{\per\square\centi\metre}. None of the velocity layers studied by \citet{ladouceur_new_2008} shows a structure coinciding with MAGIC\,2019+408. Nonetheless, the two velocity ranges showing the CO emission ($\SI{2}{\kilo\metre\per\second}<v< $\SI{10}{\kilo\metre\per\second}) and the northern HI cavity wall ($\SI{-19}{\kilo\metre\per\second}<v<\SI{-11}{\kilo\metre\per\second}$) show a sufficient HI column density at the position of MAGIC\,2019+40 (factor ~1.5 to 2 higher) even assuming an optically thin case. Regarding the different velocity ranges \citet{ladouceur_new_2008} noted that the complex motion of the gas layers in this region complicates the assignment of a velocities. The search for a counterpart will thus require a dedicated study, though currently the observation suggest a sufficient amount of target material.

One point to stress is that the presence of a dense cloud engulfed by a shock could result in a harder hadronic spectrum due to magnetic field amplification around the cloud border, which prevents low energy particles from penetrating the cloud. Nevertheless, this mechanism is only relevant when the density contrast between the cloud and the circumstellar medium is ${\gtrsim} 10^3$ \citep{Celli+2019a}. Otherwise, the amplification of the magnetic field around the cloud is not strong enough. The density contrast in our case is below that threshold, hence the spectrum inside the cloud should be very similar to the one inside the SNR.

\subsection{Implication for DSA} \label{sec:discussion}

Our interpretation of the $\gamma$-ray spectra has interesting implications for the shock acceleration theory. First, the slope of accelerated particles is constrained to be close to 4, in agreement with the standard prediction of DSA. Even a value as steep as 4.2, required by direct CR observations \cite[see, e.g.][]{evoli_galactic_2019}, is still compatible with the data.\footnote{It is important to note that strictly only the average spectrum of CR injected by SNRs needs to agree with the Galactic CR spectrum and the spectrum of a single SNR may well differ.}

In addition, we have shown that the maximum energy accelerated at the current SNR age is around a few TeV, while the maximum energy reached at the beginning of the ST phase is at most few hundreds of TeV. Hence $\gamma$-Cygni is not a PeVatron during the ST phase, nevertheless we cannot exclude that higher energies could have been reached during the very early stage of its evolution. The fact that the maximum energy now is of the order of TeV immediately provides an estimate of the level of magnetic turbulence in the shock region. Using Eq. (\ref{eq:Dout_approx}) we can estimate the diffusion coefficient upstream of the shock from the maximum energy  and the remnant age. In quasi-linear theory, the diffusion coefficient $D$ is expressed as a function of the turbulence level as  $D_{1}(p) = D_{\rm Bohm}/\mathcal{F}(k_{\rm res})$, where $D_{\rm Bohm}= r_L v/3$, $r_L$ being the Larmor radius, and $\mathcal{F}= (\delta B(k_{\rm res})/B_0)^2$ is calculated at the resonant scale $k_{\rm res}=1/r_L(E_{\max})$. Hence $E_{\max}(t_{\rm SNR})= 1$ TeV (obtained from the condition $t_{\rm acc}=t_{\rm SNR}$) inferred from the spectrum implies $\delta B/B_0\,{\simeq}\,0.25$ where we assumed $B_0= 3\mu$G. 
At a first glance, such a small value for the magnetic turbulence seems at odds with the common assumption that close to strong shocks the Bohm limit is reached. Nevertheless, two different arguments suggest that $\delta B/B_0$ should be smaller than unity as inferred here. First of all, the main mechanism often invoked to excite magnetic turbulence resonating with accelerated particles is the {\it resonant streaming instability}, which, however, saturates at $\delta B/B_0\,{\lesssim}\,1$ \cite[see][Sect. 4.2]{Blasi2013}. A higher level of turbulence requires other mechanisms, like the {\it non-resonant instability} \cite[NRI,][]{Bell2004} but such a mechanism is thought to be effective when the shock speed is ${\gg}\,1000$ km/s \cite[]{Amato-Blasi2009}. Moreover, the NRI excites small scale modes which, to efficiently scatter high energy particles, requires an inverse cascade up to the larger scales which resonate with such particles. Hence, either the NRI is not efficiently excited or it is excited but the inverse cascade does not occur on the required timescale. CRs can amplify the magnetic field upstream even through another mechanism which takes place when the medium is inhomogeneous \citep{drury_turbulent_2012}. Nevertheless, such a mechanism also requires high shock speeds to be effective. The exact condition is $u_{sh}\,{\gg}\,\sqrt{4\pi/\xi_{CR}} v_A (\delta \rho/\rho)^{-1}$, where $\delta \rho$ is the typical level of density fluctuation in the upstream while $v_A$ is the Alfv\'en speed. Using the parameter values adopted in this work, one easily gets the condition $u_{sh}\gg1000$ km/s.
Furthermore, even if the amplification were efficient, the upstream plasma may be partially neutral and the ion-neutral friction could efficiently damp the magnetic turbulence resulting in $\delta B/B_0\,{\ll}\,1$ \cite[see, e.g.,][]{nava_non-linear_2016}.

Another important piece of information inferred from the data is the time dependence of the maximum energy. As discussed in \cite{Celli+2019a} an approximate way to infer the value of $\delta$ is by equating the acceleration time with the age of the remnant. Using Eq. (\ref{eq:t_acc}) (i.e. $t_{\rm acc}(p) \propto D(p)/u^2_{\rm sh}$) and writing the diffusion coefficient again in terms of the magnetic turbulence ($D= D_{\rm Bohm} \mathcal{F}^{-1}$), we can write
 $p_{\max}(t)  \propto \mathcal{F}(t) \, u_{\rm sh}^2(t) \, t $.
In the absence of any magnetic field amplification and with a constant turbulence, the time dependence is only determined by the shock speed which is $u_{\rm sh}\,{\propto}\,t^{-3/5}$ in the ST phase, resulting in $p_{\max}(t)\,{\propto}\,t^{-1/5}$. On the contrary, our inferred value of $\delta\,{\simeq}\,2.55$ requires that the magnetic turbulence should decrease in time as $\mathcal{F}\,{\propto}\,t^{-2.35}\,{\propto}\, {u_{\rm sh}(t)}^{3.9}$. Hence, even accounting for all the uncertainties, a constant value of magnetic turbulence in the shock precursor would be difficult to reconcile with our finding which requires, instead, some level of magnetic amplification and/or damping. 
Interestingly, $\delta\,{\simeq}\,2.55$ is in good agreement with the phenomenological estimate of $\delta\,{\simeq}\,2.48$ by \citet{Gabici+2009} derived from $p_{\rm max}\,{\propto}\,t^{-\delta}$ and assuming $p_{\rm max}$($t$=200\,yr)=5\,PeV, and $p_{\rm max}$($t=5\times10^{4}$\,yr)=1\,GeV.

Concerning the propagation of escaping particles, the highest energy points detected by MAGIC clearly require a diffusion coefficient ${\sim}$ 10\,--\,35 times smaller than the average Galactic one. This ratio is a factor \numrange{2}{3e2} smaller than the prediction using Eq. (\ref{eq:Dout_approx}). However, as already discussed in Sect. \ref{sec:escaping}, Eq. (\ref{eq:Dout_approx}) only provides a lower limit on $D_{\rm out}$ and thus the two estimates can be considered compatible.
Our finding that $D_{\rm out}\,{\ll}\,D_{\rm Gal}$ is not surprising given that $D_{\rm Gal}$ (obtained from the measurements of B/C in the local CR spectrum) represents an average over the large volume of the Galactic magnetic halo and could be completely different from the diffusion coefficient in the vicinity of $\gamma$-Cygni. 

In the first place, an effective smaller diffusion coefficient could just result from the particle diffusion along magnetic flux tubes, hence 1D, rather than isotropic 3D diffusion \citep{nava_anisotropic_2013}. A second possibility is stronger turbulence around the SNR which is unsurprising due to the complexity of the Cygnus region, which counts several potential sources of turbulence (SNRs and winds from massive stars and clusters).
Alternatively, enhanced turbulence could be also naturally produced by the streaming instability of escaping CRs. Such a scenario has been investigated by several authors \citep{malkov_analytic_2013, nava_non-linear_2016, dangelo_grammage_2016} who showed that the diffusion can be easily suppressed by 1\,--\,3 order of magnitudes up to several parsecs from the SNR. Indeed, the reduced $D_{\rm out}$ of our model compared to $D_{\rm Gal}$ is compatible with the diffusion coefficient derived for the vicinity of other SNRs such as W28 \citep[][and references therein]{hanabata_detailed_2014} and W44 \citep{uchiyama_fermi_2012}.
Distinguishing between these three possibilities (1D diffusion, larger external turbulence or self-generated one) is not easy but the latter case has the advantage of being more predictive. In fact, if the escaping flux is known, the diffusion coefficient can be calculated without adding any new free parameters. 
Also note that we assumed for $D_{\rm out}$ the same Kolmogorov scaling inferred for $D_{\rm Gal}$\footnote{This is true for rigidities $\gtrsim 200$ GV, while at lower energies the CR spectrum suggests a different scaling \citep{Blasi-Amato-Serpico2012}.}. In case of turbulence coming from external sources, the Kolmogorov scaling is indeed expected, while if the turbulence is self-generated, the resulting scaling is usually different. Hence a self-consistent calculation is needed to provide a more detailed answer.

Clearly our model suffers of some limitations, mainly related to the assumptions of spherical geometry and homogeneity of the circumstellar medium. Indeed, the $\gamma$-ray map shows a patchy structure suggesting the presence of a clumpy medium. Small dense clumps may significantly modify the hadronic $\gamma$-ray spectrum as a result of magnetic field amplification occurring in the shock-cloud interaction which, in turns, modifies the propagation properties of the plasma \citep{Gabici-Aharonian2014,Inoue+2012,Celli+2019a}. Nevertheless, such an effect is mainly important at high shock speed and for a large density contrast between the clumps and the average circumstellar medium. It has been applied e.g. to the SNR RX J1713 whose shock has a speed $\sim 5000$ km/s and where the estimated density contrast is above $10^3$. In the case of $\gamma$-Cygni both quantities are much smaller, it is hence probable that such effect plays a minor role.

Another possible issue is related to contribution to the $\gamma$-ray spectrum coming from CR-illuminated clouds located along the line of sight and erroneously attributed to the SNR interior. Even if such a scenario remains possible, the CO maps do not show any clear evidence of clouds along the same line of sight of the SNR disc \citep{leung_complete_1992}, hence we are inclined to attribute all the emission from the disc region to the SNR interior.

   \section{Summary and future prospect}	    \label{sec:summary}

Combining MAGIC and {\it Fermi}-LAT data, we investigated the $\gamma$-ray emission in the vicinity of the $\gamma$-Cygni SNR, G78.2+2.1, and identified three main source components: the SNR interior, an extended emission located immediately outside the SNR that we call the arc, and a Gaussian-shaped extended source MAGIC\,J2019+408 in the north-west of the remnant. The brightness ratio between the source components is energy dependent with the SNR shell dominating below $\sim60$\,GeV and the arc only being observed by MAGIC above 250\,GeV. The morphologies and spectra of MAGIC\,J2019+408 and the arc suggest an association with the SNR. The indices for a power-law model for MAGIC\,J2019+408 and the shell are ${\sim}\,2$ in the {\it Fermi}-LAT energy range and ${\sim}\,2.8$ for the three components in the MAGIC energy range (similar within uncertainties, but explicitly: 2.55 for the shell, 2.81 for the MAGIC\,J2019+408, and 3.02 for the arc).

We interpret the three components as the result of CRs escaping the shock of the SNR upstream into the ISM, while the shock is still capable of confining less energetic CRs. In this context, the differences between MAGIC\,J2019+408 and the arc can be understood by the presence of an over-dense cloud partially engulfed by the SNR. The observed extension of the arc does not agree with the predicted size of a shock precursor based on the SNR's characteristics. 

We further presented a theoretical model to interpret the data in the framework of DSA with the inclusion of time-dependent particle escape from the SNR interior. While a leptonic origin for the $\gamma$-ray emission cannot be ruled out, given the morphologies and spectra of the source components a hadronic scenario seems more plausible. Hadronic collisions can well account for the $\gamma$-ray emission from all the three regions (the SNR's interior, the arc, and MAGIC\,J2019+408) provided that:
\begin{itemize}
   \item the spectrum of accelerated particles is $\propto p^{-4}$ and the acceleration efficiency is $\simeq 3.8\%$;
   \item the maximum energy of accelerated particles decreases in time like $\propto t^{-2.55}$ and, at the present moment, it is around few TeV while it reached an absolute maximum of about 100\,TeV at the end of the free-expansion phase;
   \item the level of magnetic turbulence in the shock region at the present moment is $\delta B/B_0 \lesssim 1$ and has to decrease in time, pointing toward the presence of self-exited magnetic waves from accelerated particles;
   \item the diffusion coefficient in the region immediately outside the SNR has to be $\sim 10-35$ times smaller than the average Galactic one as inferred from the B/C ratio in the local CR spectrum;
   \item the region around the SNR has to be patchy with extended clouds whose density is between 5 and 200 times larger than the average circumstellar medium.
\end{itemize}
All these findings agree well with the standard DSA applied to a middle-age SNR produced by a core-collapse explosion, except the quite steep time dependence of the maximum energy: the theory invoking resonant and non-resonant instabilities usually predicts flatter dependences. A way to explain the steep dependence is through some damping mechanism of the magnetic waves. Nevertheless, we stress that the description of the particle escape is not completely understood yet, even when damping processes are neglected.

Another important point of our modelling is the fact that the high energy emission from $\gamma$-Cygni needs the contribution coming from particles which have formally escaped from the acceleration process but are still diffusing around and inside the SNR. This could be of general validity for middle-aged SNRs, especially if the diffusion coefficient around those objects is suppressed with respect to the Galactic value.

The understanding of the $\gamma$-Cygni SNR could profit from future observation particularly of the molecular material, the uncovered hard X-ray to sub-GeV energy range, and very-high-energy $\gamma$-rays. Improved knowledge about the target material can reduce the uncertainty of our proposed model and determine the contribution from hadrons. Given hard X-rays or MeV $\gamma$-rays can resolve the SNR shell under the background from the pulsar, they may clarify the remaining uncertainties about a hadronic or leptonic origin by searching for possible bremsstrahlung, a cooling break in the spectrum of the SNR shell, or a pion cut-off. To validate our proposed model future $\gamma$-ray studies could test whether the extension of the arc region is energy dependent. Hence, $\gamma$-Cygni might be a prime target to study the particle escape process from SNRs particularly for deeper $\gamma$-ray observations with improved angular resolution.

   \begin{appendix}
\section{Systematic Uncertainties}
\label{sec:Sys_Uncertainties}

For both telescopes, MAGIC and {\it Fermi}-LAT, we consider two kinds of systematic uncertainties, the ones coming from the detector itself and the ones coming from the ingredients of the models supplied to the likelihood analysis.

For MAGIC \citet{aleksic_major_2016} studied the systematic uncertainties on the spectral parameters (flux normalisation, spectral index, and energy scale) resulting from the telescope performance\footnote{E.g. from instrumental uncertainties, Monte-Carlo data mismatch, and uncertainties from the analysis pipeline} for low and medium Zenith angles. Since the observations for this work were performed at the same elevations, we scaled those uncertainties with the signal-to-background ratio of each source component as described in \citet{aleksic_major_2016}. Furthermore, the authors investigated the systematic uncertainty on the source position, which we considered for the localisation of MAGIC\,J2019+408. However, \citet{aleksic_major_2016} only examined point sources or slightly extended sources. Hence, we additionally studied the uncertainties arising from the imperfect knowledge of our 2D background and exposure shape, which are part of our SkyPrism model. To estimate their effect we ran the analysis using 50 random representations of the background model and exposure map. For the former we assumed that the content in each bin of the background acceptance model in camera coordinates follows a Poisson statistics with the original value as the mean value. According to the distribution in each bins we simulate random camera background models and proceed in the same way as for the original model following \citep{vovk_spatial_2018}. For the exposure the random representations are based on the parameter uncertainties of the $\gamma$-ray acceptance model fitted to the Monte-Carlo (MC) based acceptance as described in \citep{vovk_spatial_2018}.

Due to the large energy range, the uncertainties affecting the spectral parameters are less of a concern for the morphological study. Thus, our estimates only include the uncertainties from the background and exposure model together with the effect of a possible underestimation of the MC based PSF model. \citet{aleksic_major_2016} and \citep{vovk_spatial_2018} found that the MC based PSF might be smaller by $0\fdg02$ compared to the extension of point sources. For the spectra we considered the uncertainties from telescope performance and the uncertainties of the background and exposure model. Table \ref{tab:sys_unc_spectra} contains the effect on each spectral parameter and source component for the various uncertainty origins. The uncertainties based on \citet{aleksic_major_2016} are listed under "instrument" and the background and exposure model uncertainties as "bgr+exposure".

\begin{table*}
	\caption{
     Systematic uncertainties for the spectral analysis of the {\it Fermi}-LAT and MAGIC data and each source component. The uncertainties are sorted according to their origin from either the IRFs of the detectors or the uncertainties in the shape of the background model and exposure for MAGIC and the IEMs for {\it Fermi}-LAT.}      
   \label{tab:sys_unc_spectra}
   \center
   \linespread{1.15}\selectfont
   \begin{tabular}{l c c c c c c c c}
      \hline\hline
      \multicolumn{1}{l}{Source name} & \multicolumn{5}{c}{MAGIC}\\    

      						& \multicolumn{2}{c}{$N_{0}$ [$\mathrm{TeV}^{-1}\,\mathrm{cm}^{-2}\,\mathrm{s}^{-1}$]} & \multicolumn{2}{c}{$\Gamma$} & \multicolumn{1}{c}{$E_{0}$} \\
      						& Instrument 	& Bgr + Exposure 	& Instrument 	& Bgr + Exposure 	& Instrument \\
      \hline
      SNR Shell 			& $1.6\,\times\,10^{-13}$ & $\left(\substack{+3.9 \\ -0.7}\right)\,\times\,10^{-13}$ 	& 0.21 	& $\substack{+0.22 \\ -0.13}$ 	& 15\,\% \\
      MAGIC\,J2019+408   	& $1.2\,\times\,10^{-13}$ & $\left(\substack{+1.3 \\ -0.7}\right)\,\times\,10^{-13}$ 	& 0.17 	& $\substack{+0.13 \\ -0.09}$ 	& 15\,\% \\
      Arc (annular sector)	& $0.5\,\times\,10^{-13}$ & $\left(\substack{+0.8 \\ -0.5}\right)\,\times\,10^{-13}$	& 0.16 	& $\substack{+0.15 \\ -0.12}$	& 15\,\% \\
      Arc (Gaussian model)	& $0.8\,\times\,10^{-13}$ & $\left(\substack{+0.8 \\ -0.5}\right)\,\times\,10^{-13}$	& 0.18 	& $\substack{+0.14 \\ -0.13}$	& 15\,\% \\
      \hline
      &	\multicolumn{5}{c}{{\it Fermi}-LAT}\\
      & & IEMs & & IEMs & \\
      \hline
      SNR Shell 			& $1.7\,\times\,10^{-10}$ 	& $0.1\,\times\,10^{-10}$ 	& 0.01 	& 0.001 	& 5\,\% \\
      MAGIC\,J2019+408   	& $\left(\substack{+0.6 \\ -0.5}\right)\,\times\,10^{-10}$ 	& $0.01\,\times\,10^{-10}$ 	& 0.01 	& 0.001 	& 5\,\%  \\   
      \hline
   \end{tabular}\\  
   \linespread{1.0}\selectfont
\end{table*}

For {\it Fermi}-LAT data we studied the uncertainty of the source localisation by fitting the localisation of all point sources in the RoI with associations at radio wavelength (as catalogues we used: \citep{manchester_australia_2005}, \citep{bennett_mit_1986}, \citep{douglas_texas_1996}). For source in this sample, we estimate the additional systematic uncertainty on top of the average statistical localisation uncertainty based on a Rayleigh distribution to accommodate the 68\%-percentile off-set from their catalogue position. For the estimation of the source extension we take into account the uncertainty of the PSF and the interstellar emission model (IEM). We evaluated the systematic uncertainty using the P8R2 version of the alternative IEMs from \citet{acero_first_2016} together with a $\pm$5\% scaling of the PSF. The uncertainties of the flux normalisation and spectral index result from the precision of the IEM and the effective area, so we computed it using the alternative IEMs and considering a $\pm$5\% uncertainty on the effective area (listed as instrumental uncertainty in Table \ref{tab:sys_unc_spectra}).\footnote{\url{https://fermi.gsfc.nasa.gov/ssc/data/analysis/LAT_caveats.html}}. The instrumental uncertainty on the energy scale is based on \citet{ackermann_fermi_2012}. The effects onto the spectral parameters are listed in Table \ref{tab:sys_unc_spectra}.

\end{appendix} 

\begin{acknowledgements}
We would like to thank the Instituto de Astrof\'{\i}sica de Canarias for the excellent working conditions at the Observatorio del Roque de los Muchachos in La Palma. The financial support of the German BMBF and MPG; the Italian INFN and INAF; the Swiss National Fund SNF; the ERDF under the Spanish MINECO (FPA2017-87859-P, FPA2017-85668-P, FPA2017-82729-C6-2-R, FPA2017-82729-C6-6-R, FPA2017-82729-C6-5-R, AYA2015-71042-P, AYA2016-76012-C3-1-P, ESP2017-87055-C2-2-P, FPA2017‐90566‐REDC); the Indian Department of Atomic Energy; the Japanese ICRR, the University of Tokyo, JSPS, and MEXT;  the Bulgarian Ministry of Education and Science, National RI Roadmap Project DO1-268/16.12.2019 and the Academy of Finland grant nr. 320045 is gratefully acknowledged. This work was also supported by the Spanish Centro de Excelencia ``Severo Ochoa'' SEV-2016-0588 and SEV-2015-0548, the Unidad de Excelencia ``Mar\'{\i}a de Maeztu'' MDM-2014-0369 and the "la Caixa" Foundation (fellowship LCF/BQ/PI18/11630012), by the Croatian Science Foundation (HrZZ) Project IP-2016-06-9782 and the University of Rijeka Project 13.12.1.3.02, by the DFG Collaborative Research Centers SFB823/C4 and SFB876/C3, the Polish National Research Centre grant UMO-2016/22/M/ST9/00382 and by the Brazilian MCTIC, CNPq and FAPERJ.
\newline

The \textit{Fermi} LAT Collaboration acknowledges generous ongoing support from a number of agencies and institutes that have supported both the development and the operation of the LAT as well as scientific data analysis. These include the National Aeronautics and Space Administration and the Department of Energy in the United States, the Commissariat \`a l'Energie Atomique and the Centre National de la Recherche Scientifique / Institut National de Physique Nucl\'eaire et de Physique des Particules in France, the Agenzia Spaziale Italiana and the Istituto Nazionale di Fisica Nucleare in Italy, the Ministry of Education, Culture, Sports, Science and Technology (MEXT), High Energy Accelerator Research Organization (KEK) and Japan Aerospace Exploration Agency (JAXA) in Japan, and the K.~A.~Wallenberg Foundation, the Swedish Research Council and the Swedish National Space Board in Sweden. 
Additional support for science analysis during the operations phase is gratefully acknowledged from the Istituto Nazionale di Astrofisica in Italy and the Centre National d'\'Etudes Spatiales in France. This work performed in part under DOE Contract DE-AC02-76SF00515.
\newline
This research made use of Astropy,\footnote{http://www.astropy.org} a community-developed core Python package for Astronomy \citep{robitaille_astropy_2013, astropy_astropy_2018}. 
\end{acknowledgements}

\bibliography{GCygni_GeVTeV_Morph_MAGIC_Fermi.bib}

\end{document}